\documentclass{tlp}

\usepackage{amsmath}
\usepackage{graphicx}
\usepackage{multirow}
\usepackage{listings}
\usepackage{mathtools}
\usepackage{amssymb}
\usepackage{amsmath}
\usepackage{hyperref}

\usepackage{todonotes}
\usepackage{booktabs}
\usepackage{multirow}
\usepackage{xspace}
\usepackage{tikz}
\usepackage{pgfplots}
\usepackage{subcaption} 
\usepackage{tkz-graph}
\usepackage{hhline}

\newcommand{\scaleplot}{0.7} 

\DeclareMathOperator*{\argmax}{arg\,max}

\newcommand{\added}[1]{\textcolor{black}{#1}}

\newcommand{\pf}{probabilistic fact}
\newcommand{\pfs}{probabilistic facts}
\newcommand{\wc}{weak constraint}
\newcommand{\wcs}{weak constraints}
\newcommand{\as}{answer set}
\newcommand{\ass}{answer sets}

\newcommand{\impl}{\ {:\!-}\  }
\newcommand{\weakarr}{:\sim}
\newcommand{\ASPQ}{ASP(Q)\xspace}
\newcommand{\fix}[2]{\mathit{fix_{#1}(#2)}}

\DeclareMathOperator{\optPASP}{\mathcal{P}^{\sim}}
\DeclareMathOperator{\PASP}{\mathcal{P}}
\DeclareMathOperator{\ASP}{\mathcal{T}}
\newcommand{\pasprules}{\ASP}
\DeclareMathOperator{\probfacts}{\mathcal{F}}
\newcommand{\paspfacts}{\probfacts}
\DeclareMathOperator{\weakconstraints}{\mathcal{W}}
\DeclareMathOperator{\cost}{\mathcal{C}}

\DeclareMathOperator{\world}{\mathit{w}}
\DeclareMathOperator{\worldset}{\Sigma}
\DeclareMathOperator{\selection}{\sigma}
\DeclareMathOperator{\selectionset}{\mathcal{S}}
\DeclareMathOperator{\AS}{\mathit{AS}}
\DeclareMathOperator{\optAS}{\mathit{AS}^*}
\newcommand{\base}{\mathcal{B}_{P}}

\DeclareMathOperator{\prob}{P}
\DeclareMathOperator{\lowerprob}{\underline{P}}
\DeclareMathOperator{\upperprob}{\overline{P}}
\DeclareMathOperator{\uppermpe}{\overline{MPE}}
\DeclareMathOperator{\lowermpe}{\underline{MPE}}

\newcommand{\NP}{\mathsf{NP}}

\newcommand{\PP}{\mathsf{PP}}
\newcommand{\stp}{\mathsf{\Sigma_2^P}}
\newcommand{\dtp}{\mathsf{\Delta_2^P}}

\newtheorem{definition}{Definition}
\newtheorem{example}{Example}
\newtheorem{observation}{Observation}
\newtheorem{theorem}{Theorem}

\def\naf{\ensuremath{not\ }\xspace}

\begin{document}

\lefttitle{Cambridge Author}

\jnlPage{1}{8}
\jnlDoiYr{2021}
\doival{10.1017/xxxxx}

\title[Optimal Probabilistic Answer Set Programming: Complexity and Implementations]{Optimal Probabilistic Answer Set Programming: Complexity and Implementations\thanks{DA and FR are members of the Gruppo Nazionale Calcolo Scientifico -- Istituto Nazionale di Alta Matematica (GNCS-INdAM).
This work has been partially funded by the Italian Ministry of Industrial Development (MISE) under project EI-TWIN n. F/310168/05/X56 CUP B29J24000680005 and also partially by projects  SIGENERA (CUP J29I24001770005) and MOZART (J49I24001740005) selected within the framework of the PR FESR – FSE Calabria 2021/2027 and implemented with the support of the Italian State and the Calabria Region.}}

\begin{authgrp}
\author{\gn{Damiano} \sn{Azzolini}}
\affiliation{University of Ferrara}
\author{\gn{Giuseppe} \sn{Mazzotta}}
\affiliation{University of Calabria}
\author{\gn{Francesco} \sn{Ricca}}
\affiliation{University of Calabria}
\author{\gn{Fabrizio} \sn{Riguzzi}}
\affiliation{University of Ferrara}
\end{authgrp}

\history{\sub{xx xx xxxx;} \rev{xx xx xxxx;} \acc{xx xx xxxx}}

\maketitle

\begin{abstract}
Probabilistic Answer Set Programming (PASP) extends Answer Set Programming with the possibility of representing uncertain data.
Among the possible available semantics, we focus here on the credal semantics.
In this setting, prior work has focused exclusively on plain ASP programs, i.e., programs without optimization constructs.
However, many real world tasks define optimality criteria over the possible solutions, which cannot be conveniently modeled with existing probabilistic ASP frameworks.
In this paper, we close this gap and propose optimal Probabilistic Answer Set Programming, which extends PASP by also allowing weak constraints, which are the standard optimization construct of ASP.
We define the semantics of such programs, study the computational complexity of the inference and most probable explanation tasks, and propose practical implementations based on Algebraic Model Counting and Answer Set Programming with Quantifiers.
An empirical evaluation of the proposed approach on three real world tasks confirms the applicability of our framework.
\end{abstract}

\begin{keywords}
Probabilistic Answer Set Programming, Uncertainty, Weak Constraints, Answer Set Programming with Quantifiers
\end{keywords}

\section{Introduction}
\label{sec:intro}

Answer Set Programming (ASP) is a declarative paradigm for knowledge representation and reasoning that has been applied to many crucial tasks~\citep{erdem2016applications,DBLP:journals/ki/FalknerFSTT18}, such as resource placement~\citep{azzolini2024continuous}, product configuration~\citep{DBLP:conf/lpnmr/BaumeisterHORRSW24}, software dependency resolution~\citep{DBLP:conf/sc/GamblinCBS22}, explainable artificial intelligence~\citep{kaminski2024explaining}, scheduling in health-related domains ~\citep{dodaro2026asp,dodaro2017nurse,dodaro2021asp,cappanera2023logic}, shift design~\citep{abseher2016shift}, mass spectrum analysis~\citep{kuchenmeister2024towards}, among others.

Despite its widespread success, standard ASP does not natively support probabilistic reasoning, which is often essential for modeling uncertainty in real-world scenarios~\citep{DBLP:conf/ijcai/RaedtKT07,cozman2020pasp}.
To account for this, Probabilistic ASP (PASP) extends ASP with the support of primitives that represent uncertain data, either in the form of weights~\citep{lee_talsania_wang_2017} or probabilities~\citep{cozman2020pasp,baral2009probabilistic,totis_de_raedt_kimmig_2023}.
More in detail, we adopt the PASP framework, which extends ASP with \pfs{}~\citep{DBLP:conf/ijcai/RaedtKT07}, i.e., facts with an associated probability, and adopt the credal semantics (CS)~\citep{cozman2020complexity}.
Among the key tasks of PASP, are \textit{inference}, and \textit{most probable explanation} (MPE).
%
Inference consists of computing the probability of a ground atom (the \textit{query}). Under the credal semantics (CS), each possible combination of \pfs{} defines a selection, and each selection induces a world represented as an answer set program, thereby determining the probability of that world. The probability of a query is, then, characterized by a lower and an upper bound, depending on whether the query holds in at least one or in all answer sets of the considered worlds.
On the other hand, MPE requires identifying the selection with maximum probability such that the induced world bravely or cautiously entails a given atom (often referred to as evidence).

These tasks have been extensively studied in the literature, typically in the setting of plain ASP programs, i.e., without optimization constructs. However, many real-world applications involve optimality criteria over solutions~\citep{DBLP:journals/ai/CalimeriGMR16,erdem2016applications}, which are not naturally captured by existing probabilistic ASP frameworks.

We address this gap by introducing \textit{optimal} PASP, enabling probabilistic reasoning over ASP programs with \textit{weak constraints}~\citep{DBLP:journals/tkde/BuccafurriLR00}, extended with \pfs{}.
Weak constraints are the standard ASP construct used to represent desiderata (i.e., optimization criteria) over answer sets. In particular, they select those answer sets that minimize violations, referred to as optimal answer sets.
Building on standard PASP semantics and the established semantics of weak constraints, we obtain a natural extension in which the probabilistic reasoning tasks are computed by considering only optimal answer sets.

In the paper, we define formally the semantics of PASP programs with weak constraints, study the computational complexity of the inference and most probable explanation tasks, demonstrate its modeling capabilities in encoding real-world–inspired problems, and propose practical implementations based on Algebraic Model Counting~\citep{10.1016/j.jal.2016.11.031} and Answer Set Programming with Quantifiers~\citep{DBLP:journals/tplp/AmendolaRT19}.
We also report on the results of 
an empirical evaluation of the proposed approach conducted on three interesting problems that confirm the applicability of our framework.
This paper extends the work presented in~\citep{azzolini2025novel} by refining the computational complexity proofs and providing practical implementations.

\medskip

The paper is structured as follows: after introducing the required background (Section~\ref{sec:background}), we precisely define the syntax and semantics of our proposal (Section~\ref{sec:opt_pasp}), illustrate its modeling capabilities with three real-world examples (Section~\ref{sec:modelling_examples}), provide a computational complexity study of the inference and MPE tasks (Section~\ref{sec:computational_complexity}), and propose (Section~\ref{sec:implementation} and Section~\ref{sec:experiments})  practical implementations to solve them.
Before concluding the paper (Section~\ref{sec:conclusion}), we provide a broad overview of related work (Section~\ref{sec:related}).

\section{Background}
\label{sec:background}

In this section, we recall the background involving Answer Set Programming, Probabilistic Answer Set Programming, Answer Set Programming with quantifiers, and Algebraic Model Counting. 

\subsection{Answer Set Programming}
\label{subsec:asp}

In ASP a \textit{variable} is a string starting with an uppercase letter, while a constant is a number or a string starting with a lowercase letter.
In ASP a \textit{term} is either a variable or a constant.
A \textit{standard atom} is an expression of the form $p(t_1,\ldots,t_n)$, where $p$ is \textit{predicate} of arity (i.e., number of arguments) $n$ and $t_1,\ldots,t_n$ are terms.
A \textit{standard literal} is either a standard atom $a$ or its negation $\naf a$, where $\naf$ denotes \textit{negation as failure}.
A standard literal is \textit{ground} if it contains no variables.
Let $l$ be a standard literal then $\overline{l}$ denote the \textit{complement} of $l$ that is $\overline{l} = a$ if $l = \naf a$; otherwise $\overline{l} = \naf a$.
A \textit{symbolic set} is an expression of the form $V:Conj$ where $V$ is a list of terms and $Conj$ is a conjunction of standard literals.
A \textit{ground set} is an expression of the form $v_1:conj_1;\ldots;v_n:conj_n$ where for each $i \in \{1,\ldots,n\}$, $v_i$ is a list of constants and $conj_i$ is a conjunction of ground standard literals.
An \textit{aggregate atom} is an expression of the form $\#f\{S\}\succ T$, where $f$ is an aggregate function and we consider only two function types, $\mathit{count}$ and $\mathit{sum}$, $S$ is a symbolic or a ground set, $\succ \in \{>,\geq,<,\leq\}$ is a comparison operator, and $T$ is a term called \textit{guard}.
An aggregate atom $\#f\{S\} \succ T$ is ground if $S$ is a ground set and $T$ does not contain variables.

An \textit{atom} is either a standard atom or an aggregate one.
A \textit{literal} is either an atom $a$ or its negation $\naf a$.
A literal is said to be \textit{positive} if it is an atom $a$, otherwise it is \textit{negative}.
Given a set of literals $S$, $S^+$ and $S^-$ represent the set of positive and negative literals in $S$, respectively. 

A \textit{rule} is an expression of the form 
$$h_1;\ldots;h_m\impl l_1,\ldots,l_n$$
where $n,m\geq 0$, $h_1,\ldots,h_m$ is a disjunction of standard atoms called \textit{head}, and $l_1,\ldots,l_n$ is a conjunction of literals called \textit{body}.
A \textit{weak constraint} is an expression of the form $\weakarr l_1,\ldots,l_n\ [w@l,t]$, where $n\geq 1$, $l_1,\ldots,l_n$ is a conjunction of literals called \textit{body}, $w$ and $l$ are numbers representing the \textit{weight} and the \textit{level}, respectively, and $t$ is a list of terms that can be also omitted.
In case $@l$ is omitted, the level is assumed to be $0$.

Given an ASP expression $\epsilon$ (i.e., atom, aggregate, rule, \dots), $vars(\epsilon)$ denotes the set of variables appearing in $\epsilon$, and it is ground if $vars(\epsilon)=\emptyset$.
For a rule $r$, $H_r$ is the set of atoms that appear in the head.
For a rule $r$ (resp. a \wc{} $wc$) we denote by $B_r$ (resp. $B_{wc}$) the set of literals appearing in the body.
$B_r^s$ (resp. $B_{wc}^s$) is the set of standard literals in $B_r$ (resp. $B_{wc}$) and $B_r^a$ (resp. $B_{wc}^a$) is the set of aggregate literals in $B_r$ (resp. $B_{wc}$).

A rule $r$ is said to be a \textit{fact} if $B_r = \emptyset$, a \textit{constraint} if $H_r = \emptyset$, or \textit{normal} if $\vert H_r\vert = 1$.
For a rule $r$ (resp. a \wc{} $wc$), global variables are variables that appear in $B_r^s$ (resp. $B_{wc}^s$).
Let $l$ be an aggregate literal of the form $\#f\{V:Conj\}\succ T$ appearing in the body of some rule or \wc{}.
Then, $l$ is said to be \textit{safe} if each variable in $Conj$ appears in some positive literal in $Conj$, $vars(V)\subseteq vars(Conj)$, and each variable in $T$ is global.

A rule $r$ is \textit{safe} if $vars(H_r\cup B_r^s)\subseteq vars((B_r^s)^+)$, and for each $l \in B_r^a$, $l$ is safe.
The definition of safety extends also to \wcs{}.
A \wc{} $wc$ of the form $\weakarr l_1,\ldots,l_n\ [w@l,t]$ is \textit{safe} if $vars(B_{wc}^s)\subseteq vars((B_{wc}^s)^+)$, for each $l \in B_{wc}^a$, $l$ is safe, and variables appearing in $w,l,t$ appear also in $(B_r^s)^+$.

An answer set \textit{program} is a finite set of rules and \wcs{}.
For a program $\ASP$ we denote by $\mathcal{R}(\ASP)$ and $\mathcal{W}(\ASP)$ the set of rules and \wcs{} of $\ASP$, respectively.
A program $\ASP$ is said to be \textit{plain} if $\mathcal{W}(\ASP) = \emptyset$; whereas $\ASP$ is said to be \textit{normal} if each rule in $\ASP$ is normal. 
In the following, we consider only safe programs, i.e., programs containing only safe rules and safe \wcs{}.

Given an ASP $\ASP$, we denote by $U_{\ASP}$ the set of constants appearing in $\ASP$ called \textit{Herbrand Universe}, and by $B_{\ASP}$ the set of ground standard atoms that can be obtained by predicates in $\ASP$ and constants in $U_{\ASP}$, called \textit{Herbrand Base}. 

A \textit{variable substitution} $\theta:X\mapsto U_{\ASP}$ is a total function that maps variables from the set $X$ to constants in $U_{\ASP}$.
Given an ASP expression $\epsilon$ (i.e., atom, literal, rule, \dots), $\theta(\epsilon)$ denotes the expression obtained from $\epsilon$ by substituting each occurrence of variables in $V$ with the constant they are mapped to.
Given a symbolic set $S$ of the form $V:Conj$, then $\mathit{ground}(S)$ denotes the ground set made of pairs $\theta(V):\theta(Conj)$, where $\theta:vars(Conj)\mapsto U_{\ASP}$ is a possible variable substitution.

Let $r\in \ASP$ be a rule of the form $h^1;\ldots;h^m \impl l^1,\ldots,l^n$.
An instantiation of $r$ is a rule of the form $h_1;\ldots; h_m\impl l_1,\ldots,l_n$ where $\theta:vars((B_r^s)^+)\mapsto U_{\ASP}$, for each $i \in \{1,\ldots,m\}$, $h_i = \theta(h^i)$, and for each $j \in \{1,\ldots,n\}$ if $l^j \in B_r^s$, then $l_j = \theta(l^j)$, otherwise $l_j$ is of the form $\#f\{\mathit{ground}(\theta(V:Conj))\}\succ\theta(T)$.
We denote by $\mathit{ground}(r)$ the set of all possible ground instantiations of $r$.
Let $wc\in \ASP$ be a \wc{} of the form $\weakarr l^1,\ldots,l^n\ [W@L,T]$.
An instantiation of $wc$ is a \wc{} of the form $\weakarr l_1,\ldots,l_n [w@l,t]$ where $\theta:vars((B_{wc}^s)^+)\mapsto U_{\ASP}$, for each $i \in \{1,\ldots,n\}$ if $L_i \in B_{wc}^s$ then $l_i = \theta(L_i)$, otherwise $l_i$ is of the form $\#f\{\mathit{ground}(\theta(V:Conj))\}\succ\theta(T)$, $w = \theta(W)$, $l = \theta(L)$, and $t = \theta(T)$.
For an ASP $\ASP$, $\mathit{ground}(\ASP)$ denotes the union of all possible ground instantiations for each rule and \wc{} in $\ASP$.

\begin{example}
Let $\ASP$ be the following program:
{\normalfont
\begin{lstlisting}
c(1,2). c(2,1). d(1). d(2). 
a(X,Y):- d(X), d(Y).
b(X):- c(X,Y), #count{Z:a(X,Z)} > Y.
:|$\sim$| b(X). [X@1]
\end{lstlisting}
}
The first line contains 4 facts, the second line contains a normal rule, the third line also represents a normal rule but with an aggregate in the body, while the last rule contains a \wc{}.
Then, $\mathit{ground}(\ASP)$ denotes the following ground program 
{\normalfont
\begin{lstlisting}
c(1,2). c(2,1). d(1). d(2). 
a(1,1):- d(1),d(1).
a(1,2):- d(1),d(2).
a(2,1):- d(2),d(1).
a(2,2):- d(2),d(2).
b(1):- c(1,2), #count{1:a(1,1); 2:a(1,2)} > 2.
b(2):- c(2,1), #count{2:a(2,1); 2:a(2,2)} > 1.
:|$\sim$| b(1). [1@1]
:|$\sim$| b(2). [2@1]
\end{lstlisting}
}
\end{example}
The dependency graph of $\ASP$, denoted by $G_{\ASP}$, is a directed labeled graph whose nodes are atoms appearing in $ground(\ASP)$ and there exists a positive (resp. negative) edge $(u,v,+)$ (resp. $(u,v,-)$) if there exists a rule $r \in ground(\ASP)$ such that $v$ appears in the head of $r$ and for $u$ at least one of the following conditions holds: (i) $u$ appears in $B_r^s$ (resp. $not\ u$ appears in $B_r^s$); (ii)  $B_r^a$ contains an aggregate of the form $\#f\{V:Conj\}\succ T$ such that $u$ appears in $Conj$ (resp. $not\ u$ appears in $Conj$).
$\ASP$ is said to be \textit{stratified} if $G_{\ASP}$ contains no loops involving negative edges; 
\textit{head-cycle-free} (HCF) if for each $r \in ground(\ASP)$ and each pair of atoms $u,v \in H_r$ there are no positive loops in $G_{\ASP}$ involving both $u$ and $v$.

An \textit{interpretation} $I\subseteq B_{\ASP}$ is a set of ground standard atoms.
A positive (resp. negative) ground standard literal $l = a$ (resp. $l = \naf a$) is true w.r.t. $I$, denoted by $I\models l$, if $a \in I$ (resp. $a \notin I$); otherwise, $l$ is false w.r.t. $I$, denoted by $I\not\models l$.

Let $S$ be a ground set of the form $v_1:conj_1;\ldots;v_n:conj_n$, then $\mathit{Eval}(S,I) = \{v_i\mid 1\leq i\leq n, I\models conj_i\}$.
Let $l$ be a ground aggregate literal of the form $\#f\{S\}\succ t$, then $I \models l$ iff $\sum\limits_{v_i \in \mathit{Eval}(S,I)} v_i^0 \succ t$ holds, with $v_i^0 = v_i[0]$ if $f = sum$, otherwise, $v_i^0 = 1$. 
A disjunction of ground standard literals $l_1;\ldots;l_n$ is true w.r.t. $I$, denoted by $I\models l_1;\ldots;l_n$, if $I \models l_i$, for some $i \in \{1,\ldots,n\}$.
A conjunction of ground standard literals $l_1,\ldots,l_n$ is true w.r.t. $I$, denoted by $I\models l_1,\ldots,l_n$, if $I \models l_i$, for each $i \in \{1,\ldots,n\}$.
Let $r \in \mathit{ground}(\ASP)$ be a rule, then $r$ is satisfied w.r.t. $I$ if $B_r$ is false w.r.t. $I$ or $H_r$ is true w.r.t. $I$.
An interpretation $I$ is a \textit{model} of $\ASP$ if $I$ satisfies each rule in $\mathit{ground}(\ASP)$.

Let $\ASP$ be a program and $I$ be an interpretation, then $\ASP^I$ denotes the FLP-reduct~\citep{DBLP:journals/ai/FaberPL11}, that is, the program obtained from $\ASP$ by removing each rule having a false body w.r.t. $I$.
A model $I$ is an \textit{answer set} of $\ASP$ if $I$ a $\subset$-mininal model of $\ASP^I$ (i.e., there is no $I' \subset I$ such that $I'$ is a model of $\ASP^I$).
For an ASP $\ASP$ we denote with $\AS(\ASP)$ the set of its answer sets.

For $M \in \AS(\ASP)$, \wcs{} in $\mathit{ground}(\ASP)$ define the cost of $M$.
More precisely, the set of \textit{\wc{} violations} is defined as $ws(\ASP,M) = \{w,l,t\mid \weakarr l_1,\ldots,l_n [w@l,t] \in \ASP, M \models l_1,\ldots,l_n\}$. 
Then, for each level $l$, the cost of $M$ at level $l$ is defined as $\mathcal{C}(M,\ASP,l) = \sum\limits_{w,l,t \in ws(\ASP,M)} w$.
Let $M_1,M_2 \in \AS(\ASP)$, then $M_1$ is \textit{dominated} by $M_2$ if there exists an level $l$ such that $\mathcal{C}(\ASP,M_1,l)>\mathcal{C}(\ASP,M_2,l)$, and for each $l' > l$, $\mathcal{C}(\ASP,M_1,l')=\mathcal{C}(\ASP,M_2,l')$.
An answer set $M$ is also an \textit{optimal answer set} if, for each $M' \in \AS(\ASP)$, $M$ is not dominated by $M'$.
We denote by $\optAS(\ASP)$ the set of optimal answer sets of $\ASP$.
In what follows, we consider ground head-cycle-free programs containing only stratified  aggregates~\citep{DBLP:journals/tplp/AlvianoFG23}.
Note that stratified aggregates do not introduce any additional source of complexity~\citep{DBLP:journals/tplp/AlvianoFG23}, but they are instrumental in providing natural and intuitive encoding in our proofs.

\subsection{Probabilistic Answer Set Programming}
\label{subsec:pasp}

\textit{Probabilistic facts}~\citep{DBLP:conf/ijcai/RaedtKT07} are among the simplest syntactical constructs to express uncertainty in a logic-based language.
In detail, a \pf{} has the form $\pi::a$ where $a$ is a ground standard atom and $\pi \in [0,1]$ is a probability.
A probabilistic answer set program (PASP, and we use the same acronym to also denote Probabilistic Answer Set Programming) $\PASP$ is a tuple $\PASP = (\ASP,\probfacts)$ where $\ASP$ is a plain head-cycle-free ASP and $\probfacts$ is a set of \pfs{}.
With $\probfacts^a$ we identify the ground standard atoms associated with the \pfs{} in $\probfacts$.
We impose that \pfs{} cannot appear in heads of rules.
This is a standard assumption in the literature~\citep{DBLP:conf/ijcai/RaedtKT07}.
A subset of $\probfacts^a$ identifies a \textit{selection} $\selection$.
For a PASP with $n$ \pfs{}, there are $2^n$ possible selections.
A selection $\selection$ identifies a \textit{world} $\world_{\selection}$, i.e., an answer set program obtained by adding each atom in $\sigma$ to $\ASP$.
We denote with $\selectionset$ and $\worldset$ the set of all possible selections and worlds, respectively.

To assign a meaning to PASPs, we consider the \textit{credal semantics} (CS)~\citep{cozman2020pasp}.
The probability of a world $\world_{\selection}$, $\prob(\world_{\selection})$, is computed by multiplying the probabilities of the \pfs{} according to whether they are included or not in the world $\world_{\selection}$.
More precisely,
\begin{equation}
\label{eq:prob_world}
\prob(\world_{\selection}) = \prod_{a_i \in \world_{\selection}} \pi_i \cdot \prod_{a_i \not \in \world_{\selection}} (1-\pi_i)
\end{equation}
where $\probfacts = \{\pi_1 :: a_1, \dots, \pi_n :: a_n\}$.
We also use the notation $\prob(\sigma)$ for a selection $\sigma$ and we set it to $\prob(w_\sigma)$.

The probability of a ground atom $q$ (the \textit{query}) is characterized by a lower $\lowerprob(q)$ and an upper $\upperprob(q)$ bound (also called \textit{lower} and \textit{upper} probability).
In words, a world contributes to the the upper probability if the query is present in some \ass{} while it contributes to the lower probability if the query is present in all \ass{}.
In formulas,
\begin{equation}
\label{eq:lower_prob_pasp}
\lowerprob(q) = \sum_{\world_{\selection} \in \worldset \mid \forall M \in \AS(\world_{\selection}), \ M \models q} \prob(\world_{\selection}),
\end{equation}

\begin{equation}
\label{eq:upper_prob_pasp}
\upperprob(q) = \sum_{\world_{\selection} \in \worldset \mid \exists M \in \AS(\world_{\selection}), \ M \models q} \prob(\world_{\selection}).
\end{equation}
Computing the probability of a query is called \textit{inference}.

The CS requires that every world has \textit{at least one answer} set, i.e., to be coherent.
This is necessary as it ensures that, for a query $q$, $\lowerprob(q) + \upperprob(not \ q) = 1$.
If some worlds are incoherent, then there is a loss of probability mass, and the above relation does not hold.
\added{In this paper, we always assume coherent PASP, unless otherwise stated.}

The \textit{most probable explanation}~\citep{DBLP:conf/ijcai/AzzoliniMRR25} task in PASP requires finding the selection which generates the world associated with the highest probability (computed with Equation~\ref{eq:prob_world}) where a ground literal $e$ (the \textit{evidence}) holds cautiously (i.e., in every \as{}) or bravely (i.e., in at least one \as{}).
These two are defined as $\lowermpe(e)$ and $\uppermpe(e)$, respectively, and are computed as
\begin{equation}
\label{eq:mpe_pasp_lower}
\lowermpe(e) = \argmax_{\sigma \in \selectionset \mid  \forall M \in AS(\world_{\selection}), M \models e } \prob(\world_{\selection}),
\end{equation}
\begin{equation}
\label{eq:mpe_pasp_upper}
\uppermpe(e) = \argmax_{\sigma \in \selectionset \mid \exists M \in AS(\world_{\selection}), M \models e } \prob(\world_{\selection}).
\end{equation}
We denote with $\prob(\lowermpe(e))$ (resp. $\prob(\uppermpe(e))$) the probability of the selection associated with $\lowermpe(e)$ (resp. $\uppermpe(e)$).

\begin{table}[tb]
    \centering
    \caption{Selections, probabilities, and answer sets for Example~\ref{ex:propositional_pasp}.}
    \label{tab:worlds_answer_sets_program}
    \begin{tabular}{c|c|c|c}
    \topline
        $id$ & $\sigma$ & $\prob(w_\sigma)$ & $AS(w_\sigma)$  \\
        \cline{1-4}
        $w_0$ & $\{\}$ &  $0.8 \cdot 0.7 = 0.56$ & $\{\}$ \\ 
        $w_1$ & $\{b\}$ & $0.8 \cdot 0.3 = 0.24$ & $\{\{b,q\}, \{b,nq\}\}$ \\ 
        $w_2$ & $\{a\}$ & $0.2 \cdot 0.7 = 0.14$ & $\{\{a,q\}\}$ \\ 
        $w_3$ & $\{a,b\}$ & $0.2 \cdot 0.3 = 0.06$ & $\{\{a,b,q\}\}$ \\ 
    \end{tabular}
\end{table}

\begin{example}
\label{ex:propositional_pasp}
Consider the following simple PASP, which will be our running example.
\begin{lstlisting}
0.2::a.
0.3::b.
q:- a.
q;nq:- b.
\end{lstlisting}
There are 4 worlds, whose probabilities and answer sets are listed in Table~\ref{tab:worlds_answer_sets_program}.
If we consider the query $q$, it is present in every answer set of $w_2$ and $w_3$, so these worlds contribute to the lower probability.
For the upper probability, $w_1$ contributes as well.
Overall, $\lowerprob(q) = 0.2$ and $\upperprob(q) = 0.44$.
Similarly, $\lowermpe(q) = \{a\}$ and $\uppermpe(q) = \{b\}$.
\end{example}

\subsection{Answer Set Programming with Quantifiers}
\label{subsec:aspq}
We now introduce Quantified Answer Set Programming (\ASPQ) which will be instrumental in our complexity studies.
\ASPQ extends ASP by introducing the concept of \textit{quantifiers} over (optimal) answer sets of ASP programs~\citep{DBLP:journals/tplp/AmendolaRT19,DBLP:journals/tplp/MazzottaRT24}.
With quantifiers, we can model problems throughout the entire PH as naturally as we can model $\NP$ problems in ASP.
An \ASPQ program $\Pi$ is of the form: 
\begin{equation}\label{eq:aspq_prg}
\Box_1 \ASP_1 \ldots \Box_n \ASP_n: C,
\end{equation}
where, for each $i \in \{1,\ldots,n\}$, $\ASP_i$ is a normal ASP program, $\Box_i \in \{\forall^{st},\exists^{st}\}$, and $C$ is a normal program both plain and stratified.
An \ASPQ{} program is said to be \textit{existential} if $\Box_1 = \exists$, otherwise it is \textit{universal}.
Moreover, we say that an \ASPQ program is \textit{uniform} if $\Box_i = \Box_{i+1}$, for each $i \in \{1,\ldots,n-1\}$. 
Let $\Pi$ be an \ASPQ of the form~(\ref{eq:aspq_prg}), $\ASP$ be a program, and $M \in AS(\ASP)$. 
\added{Then}, $fix_{\ASP}(M)$ denotes the set of facts and constraints $\{a \mid a \in M\}\cup\{\impl a\mid a \in B_{\ASP}\setminus M\}$ and $\Pi_{\ASP,M}$ the \ASPQ of the form $\Box_1 \ASP_1 \cup \fix{\ASP}{M} \ldots \Box_n \ASP_n:C$.
The \ASPQ semantics can now be defined inductively:
\begin{itemize}
    \item $\exists^{st}\ASP:C$ is coherent iff there exists $M \in \optAS(\ASP)$ such that $C\cup fix_{\ASP}(M)$ is coherent.
    \item $\forall^{st}\ASP:C$ is coherent iff for each $M \in \optAS(\ASP)$, $C\cup fix_{\ASP}(M)$ is coherent.
    \item $\exists^{st}\ASP \Pi$ is coherent iff there exists $M \in \optAS(\ASP)$ such that $\Pi_{\ASP,M}$ is coherent.
    \item $\forall^{st}\ASP \Pi$ is coherent iff for each $M \in \optAS(\ASP)$, $\Pi_{\ASP,M}$ is coherent.
\end{itemize}
For an existential \ASPQ program $\Pi$ of the form~\eqref{eq:aspq_prg}, $M \in \optAS(\ASP_1)$ is a \textit{quantified answer set} of $\Pi$ iff $\Box_2 \ASP_2 \cup \fix{\ASP_1}{M_1} \ldots \Box_n \ASP_n: C$ is coherent.
$QAS(\Pi)$ denotes the quantified answer sets of $\Pi$.

\subsection{Algebraic Model Counting}
\label{subsec:amc}

Algebraic Model Counting (AMC)~\citep{10.1016/j.jal.2016.11.031} is a generic framework that generalizes the well-known Weighted Model Counting (WMC) tasks to \textit{commutative semirings}~\citep{10.5555/1386688}.

\begin{definition}[Commutative Semiring]
\label{def:commutative_semiring}
A commutative semiring is a tuple $R= (D,\oplus,\otimes,0,1)$
where 
$\mathcal{D}$ is the domain of interest (for example, $\mathbb{N}$, $\mathbb{R}$, etc),
$\oplus$ and $\otimes$ are associative and commutative operations,
and 0 and 1 are the neutral elements for $\oplus$ and $\otimes$, respectively.
\end{definition}

\begin{definition}[Algebraic Model Counting]
\label{def:amc}
Given a propositional theory $T$ over variables $V$, a commutative semiring $R$, and a weight function $w : \mathit{lit}(V) \mapsto A$, where $\mathit{lit}(V)$ is the set of propositional literals build with the variables in $V$, $AMC(T)$ is defined as
$$
AMC(T) = \bigoplus_{M \in \mathit{models}(T)} \bigoplus_{l \in M} w(l)
$$
where $\mathit{models}(T)$ is the set of models of $T$.
\end{definition}
By selecting an appropriate semiring, it is possible to encode many different tasks (see~\citep{10.1016/j.jal.2016.11.031} for a list of examples).

Despite the generality of AMC, some tasks in Statistical Relational Artificial Intelligence~\citep{raedt2016statistical} can be better formalized within the Second-Level Algebraic Model Counting (2AMC) framework~\citep{DBLP:journals/tplp/KieselTK22}.
Intuitively, 2AMC extends AMC by nesting an AMC task inside another AMC task.

\begin{definition}[Second-Level Algebraic Model Counting]
\label{def:2amc}
Given a 
propositional theory $T$ whose variables are partitioned into $(V_i,V_o)$, 
two commutative semirings
$R_{i} = (D^{i},\oplus^i,\otimes^i,n_{\oplus}^i,n_{\otimes}^i)$ and
$R_{o} = (D^{o},\oplus^o,\otimes^o,n_{\oplus}^o,n_{\otimes}^o)$,
two weight functions, $w_i$ and $w_o$, and 
a transformation function $f_{io}$ mapping the values of $R^i$ to $R^o$, 
let 
$\mathcal{T} = (T,V_i,V_o,\mathcal{R}_i,\mathcal{R}_o,w_i,w_o,f_{io})$,
2AMC on $T$ is defined as: 
\begin{align}
\label{eq:2amc}
    2AMC(\mathcal{T}) = 
    \bigoplus\nolimits_{I_{o} \in \mu(V_{o})}^{o}
    \bigotimes\nolimits^{o}_{a \in I_{o}} 
    w_{o}(a) 
    \otimes^{o}
        f_{io}(
          \bigoplus\nolimits_{I_{i} \in \varphi(T \mid I_{o})}^{i} \bigotimes\nolimits^{i}_{b \in I_{i}} w_{i}(b)
        )
\end{align}
where $\mu(V_o)$ is the set of possible assignments to $V_o$ and $\varphi(T \mid I_{o})$ the set of assignments $I_i$ of $V_i$ such that $(I_i,I_o)$ satisfies $T$. 
\end{definition}

\section{Optimal Probabilistic ASP}
\label{sec:opt_pasp}

We now introduce the \textit{Optimal} Probabilistic Answer Set Programming (optimal PASP/optPASP) framework.

\begin{definition}
\label{def:opt_pasp}
An optimal PASP $\optPASP$ is a tuple $(\ASP,\probfacts,\weakconstraints)$, where 
$\ASP$ is a plain head-cycle-free ASP, 
$\probfacts$ is a set of \pfs{}, and 
$\weakconstraints$ is a set of \wcs{}.
\end{definition}
In other words,  an \textit{optimal} PASP is a PASP extended with a set of \wcs{}.
The set of \wcs{} imposes an optimality criterion on the possible answer sets for a world.
As for plain PASP (i.e., PASP without \wcs{}), we require that each world has at least one (optimal) answer set.
Clearly, if $\world_{\selection}$ has only one answer set $A$, then $A$ will also be the optimal answer set.

The probability bounds of a query $q$ in an optimal PASP are computed by checking whether query is true in the optimal answer sets of the worlds: a world contributes to the the upper probability if the query is present in some optimal answer sets while it contributes to the lower probability if the query is present in all optimal answer sets.
This is analogous to Equations~\eqref{eq:lower_prob_pasp} and~\eqref{eq:upper_prob_pasp} which target plain PASP, the only difference is in the answer sets to check (all for plain PASP while only the optimal ones for optPASP).
In formulas, given an optimal PASP $\optPASP$, the lower (resp. upper) optimal probability $\lowerprob^*(q)$ (resp. $\upperprob^*(q))$ for a query $q$ are computed as
\begin{equation}
\label{eq:lower_prob_opt_pasp}
\lowerprob^*(q) = \sum_{\world_{\selection} \in \worldset \mid \forall M \in \optAS(\world_{\selection}), \ M \models q} \prob(\world_{\selection}),
\end{equation}

and

\begin{equation}
\label{eq:upper_prob_opt_pasp}
\upperprob^*(q) = \sum_{\world_{\selection} \in \worldset \mid \exists M \in \optAS(\world_{\selection}), \ M \models q} \prob(\world_{\selection}).
\end{equation}

We now introduce the MPE task for optPASP, analogously to the definition for PASP (Equation~\ref{eq:mpe_pasp_lower} and~\ref{eq:mpe_pasp_upper}). 
Given a ground literal $e$ (the evidence), the lower (cautious) and upper (brave) optimal MPE states are now computed by considering only optimal answer sets.
That is,
\begin{equation}
\label{eq:mpe_opt_pasp_lower}
\lowermpe^*(e) = \argmax_{\sigma \in \selectionset \mid  \forall M \in \optAS(\world_{\selection}), M \models e } \prob(\world_{\selection}),
\end{equation}

\begin{equation}
\label{eq:mpe_opt_pasp_upper}
\uppermpe^*(e) = \argmax_{\sigma \in \selectionset \mid \exists M \in \optAS(\world_{\selection}), M \models e } \prob(\world_{\selection}).   
\end{equation}


The next theorem relates the probability of a query computed in a plain PASP to that computed in an optimal PASP.

\begin{theorem}
\label{th:prob_pasp_opt_pasp}
Let $\ASP$, $\probfacts$, and $\weakconstraints$ be a set of normal rules, \pfs{}, and \wcs{}, respectively, and $q$ an atom (query).
Let $\PASP = (\ASP, \probfacts)$ and $\optPASP = (\ASP, \probfacts, \weakconstraints)$.
Then, 
$\lowerprob(q) \leq \lowerprob^*(q) \leq \upperprob^*(q) \leq \upperprob(q)$.
\end{theorem}
\begin{proof}
The number of worlds of $\PASP$ and $\optPASP$ is the same, as is the probability for each world, since the set of \pfs{} is the same.
Consider a world $w$ in $\PASP$ and its counterpart $w^*$ for $\optPASP$.
Since $AS^*(w^*)\subseteq AS(w)$, if the condition for the world to contribute to the lower bound is satisfied for $\PASP$,  then it will be satisfied for $\optPASP$ but not vice versa (the condition for $\PASP$ is stronger).
Conversely, if the condition for the world to contribute to the upper bound is satisfied for $\optPASP$,  then it will be satisfied for $\PASP$ but not vice versa (the condition for $\optPASP$ is stronger).
\end{proof}

Theorem~\ref{th:prob_pasp_opt_pasp} shows that the semantics of optimal PASP is credal \added{as} it is the set of all probability measures that dominate an infinitely monotone Choquet capacity~\citep{augustin2014introduction,DBLP:journals/jair/CozmanM17}.
As in the credal semantics for plain PASP, the distribution of the probability mass of a world over the optimal answer set is left open.
The only difference is that the set of answer sets is a subset of those for plain PASP. 
Moreover, the following theorems immediately hold.
\begin{theorem}
Let $\PASP = (\ASP, \probfacts)$ with query $q$ and $\optPASP = (\ASP, \probfacts, \weakconstraints)$ with 
$\weakconstraints = \{:\sim q \ [C]\}$ with $C < 0$ or with 
$\weakconstraints = \{:\sim \overline{q} \ [C]\}$ with $C > 0$.
Then, $\lowerprob^*(q) = \upperprob^*(q) = \upperprob(q)$.
\end{theorem}
\begin{theorem}
Let $\PASP = (\ASP, \probfacts)$ with query $q$ and $\optPASP = (\ASP, \probfacts, \weakconstraints)$ with 
$\weakconstraints = \{:\sim q \ [C]\}$ with $C > 0$ or with 
$\weakconstraints = \{:\sim not \ q \ [C]\}$ with $C < 0$.
Then, $\lowerprob^*(q) = \upperprob^*(q) = \lowerprob(q)$.
\end{theorem}

\begin{observation}
\label{obs:relax_strong_constraints}
The credal semantics is not defined when there is at least one incoherent world.
Incoherencies are usually due to constraints (or odd cycles through negation). 
The possibility of using \textit{weak} constraints allows us to relax ``strong'' constraints, making such problems coherent and interpretable under the credal semantics.
\end{observation}

\begin{example}
\label{ex:propositional_optpasp}
Consider Example~\ref{ex:propositional_pasp} extended with a weak constraint and a rule, thus obtaining an optPASP:
\begin{lstlisting}
0.2::a.
0.3::b.
q:- a.
q;nq:- b.
qr:- nq.
:|$\sim$| qr.[1]
\end{lstlisting}
Here, the world $w_1$ obtained with selection $\{b\}$ (see Table~\ref{tab:worlds_answer_sets_program}) will have only one \textit{optimal} answer set, namely $\{b,q\}$.
Here, $\lowerprob^*(q) = \upperprob^*(q) = 0.44$.
Then, $\uppermpe^*(q) = \lowermpe^*(q) = \{b\}$.
If we change the last rule to $qr \impl q$, the unique optimal answer set for $w_1$ will be $\{b,nq\}$ so, in this case, $\lowerprob^*(q) = \upperprob^*(q) = 0.2$.
Now, $\uppermpe^*(q) = \lowermpe^*(q) = \{a\}$.
For this example, the lower and upper probability (resp. MPE states) coincide, but this does not hold in general.
\end{example}

\section{Modeling Examples}
\label{sec:modelling_examples}

In this section, we show how three real world problems can be naturally encoded within the optPASP framework.

\paragraph{Optimal Reviewer Assignment.}
The optimal reviewer assignment problem has received substantial interest in literature~\citep{10.1007/978-3-031-71481-8_25,Kalmukov2020,diperro2025,DBLP:conf/aiia/AmendolaDLR16}, with many different variations.
In brief, it consists of assigning reviewers to articles for peer review.
For ease of explanation, here we consider a simplified scenario where each reviewer may accept the invitation and terminate her/his reviews within the deadline with a certain probability and in which each paper may be assigned to at least one reviewer.
\added{The probabilities may derive from historical data.
For example, a particular reviewer was invited to be a member of the program committee of $N$ conferences and $X$ times he/she accepted the invitation and completed the reviews within the expected deadline.
So, the value $X/N$ can be the probability value associated to that particular reviewer.  }
The goal is to find a set of paper assignments such that the number of papers assigned to each reviewer is roughly the same.
This task can be encoded with the following $\optPASP$, for the case of \added{three} reviewers \added{and two papers}.
\begin{lstlisting}
0.6::reviewer(r1). 0.7::reviewer(r2).  0.8::reviewer(r3).

paper(p1). paper(p2).
willing(R,P):- reviewer(R), paper(P).

assign(R,P):- willing(R,P), not skip(R,P).
skip(R,P):- willing(R,P), not assign(R,P).

reviewed(P):- assign(R,P).

load(R,2):- assign(R,P1), assign(R,P2), P1!=P2.
load(R,1):- assign(R,P), not load(R,2).
load(R,0):- reviewer(R), not load(R,1), not load(R,2).
less_than_two :- load(R,C), C < 2.

delta(R0,R1,D):- load(R0,L0), load(R1,L1), R0 != R1, L0<=L1, D=L1-L0.
:|$\sim$| assign(R1,P), assign(R2,P), assign(R3,P), R1!=R2, R1!=R3, R2!=R3. [1@2,P]
:|$\sim$| paper(P), not reviewed(P). [1@1,P]
:|$\sim$| delta(R0,R1,D). [D@0,R0,R1]
\end{lstlisting}
Articles to review are identified with the predicate $\mathit{paper}/1$ while reviewers with $\mathit{reviewer}/1$ \pfs{}.
For simplicity, we assume that every reviewer is willing to review any article ($\mathit{willing}/2$) and can ($\mathit{assign}/2$) or not ($\mathit{skip}/2$) be assigned to a paper.
The predicate ($\mathit{load}$/2) computes the load of a reviewer, i.e., the number of papers that she/he has to review.
The difference in load between each pair of reviewers is represented by $\mathit{delta}/3$.
Two \wcs{} impose optimality criteria on the solutions:
\begin{itemize}
    \item the weak constraint at the priority level $2$ minimizes the number of papers assigned to more than two reviewers;
    \item the weak constraint at the priority level $1$ minimizes the number of articles not reviewed;
    \item the weak constraint at the priority level $0$ minimizes the difference in load among reviewers.
\end{itemize}
By asking $\mathit{less\_than\_two}$, we can compute the probability (bounds) that at least one reviewer is assigned with less than two papers (which, for the previous program, is $[0.336,0.788]$).
\added{We can also ask other queries within this program. 
For instance, we can compute the probability that at least one paper received exactly one review, by adding rules of the form}
\begin{lstlisting}
one_review :- assign(r1,P), not assign(r2,P), not assign(r3,P).
one_review :- assign(r2,P), not assign(r1,P), not assign(r3,P).
one_review :- assign(r3,P), not assign(r1,P), not assign(r2,P).
\end{lstlisting}
\added{
and then asking the probability of $one\_review$.
}
\added{Overall, with our proposed formalism, it is now possible to answer probability queries involving the \textit{optimal} assignment for reviewers, which were not possible with plain PASP.}


\paragraph{Scheduling Problem under No-Show Behavior.}
The patient scheduling and overbooking problem under no-show behavior~\citep{Topuz2024,cayitli2012universal,alaeddini2011probabilistic,dodaro2021asp} describes a scenario where patients should undergo different medical exams that require different resources.
As patients may not show up, the facility adopts overbooking to avoid leaving resources idle.
However, this may result in patients that actually show up, but do not have the required resource assigned: in this case, they should be refunded.
The following $\optPASP$ models the problem in a simplified scenario.
\begin{lstlisting}
0.4::show(1). 0.4::show(2). 0.4::show(3).
0.4::show(4). 0.4::show(5).

res(r1,chair). res(r3,bed).
res(r2,chair). res(r4,bed).
exam(e1,chair). exam(e2,bed).
take(1,e1). take(2,e2). take(3,e1). 
take(4,e2). take(5,e2).
require(P,R):- take(P,E),exam(E,R).

assign(P,R,Id) :- show(P), require(P,R), res(Id,R), not nAssign(P,R,Id).
nAssign(P,R,Id) :- show(P), require(P,R), res(Id,R), not assign(P,R,Id).
:- assign(P1,R,I), assign(P2,R,I), P1!=P2.

fulfilled(P):- show(P), assign(P,R,I).
two_not_fulfilled:- show(A),show(B), A!=B, not fulfilled(A), not fulfilled(B). 

refund_cost(P,2):- show(P), not fulfilled(P).
resource_cost(R,1):- assign(P1,chair,R).
resource_cost(R,2):- assign(P1,bed,R).

:|$\sim$| refund_cost(R,C). [C@1,R]
:|$\sim$| resource_cost(R,C). [C@0,R]
\end{lstlisting}
The \pfs{} of the form $\mathit{show}(X)$ with $X \in \{1,\dots,5\}$, denote the probabilities with which the patients may or may not show up.
The facts of the form $\mathit{exam}(E,R)$ state that exam $E$ requires resource $R$.
We assume that there are two exams, $\mathit{e1}$ and $\mathit{e2}$, and two types of resources, $\mathit{chair}$ and $\mathit{bed}$, each associated with an id ($\mathit{res(Id,R)}$).
The facts of the form $\mathit{take(P,E)}$ state that the patient $P$ undergoes exam $E$.
A resource of type $R$ with id $Id$ can be assigned ($\mathit{assign(P,R,Id)}$, and in this case, the request has been $\mathit{fulfilled}$) or not assigned ($\mathit{nAssign(P,R,Id})$) to a patient $P$ who showed up and actually required that resource.
The constraint prohibits assigning a resource to two patients at the same time.
The predicate $\mathit{refund\_cost(P,R)}$ states that the cost for refunding a patient $P$ that is present but not fulfilled is $R$.
Similarly, used resources are associated with a cost through $\mathit{resource\_cost}/2$.
Two \wcs{} impose optimality of the solutions: we want to minimize the cost of the refund (level 1) and then the cost of the used resources (level 0).
Here we may be interested in computing the probability of having two patients not fulfilled, by querying $\mathit{two\_not\_fulfilled}$, obtaining $[0.0256, 0.16]$.
\added{In general, with optimal PASP we can compute probabilities of events in scenarios of optimal resource allocation. }

\paragraph{Viral Marketing.}
The viral marketing problem is a well-studied problem in decision theory~\citep{DBLP:conf/aaai/BroeckTOR10,10.1145/502512.502525,azzolini2025solving}.
It considers a social network of people, represented by a graph.
Some of them can be targeted by an ad.
Friendship relations influence the shopping behavior of the people involved.
The following optimal PASP, adapted from~\citep{DBLP:conf/aaai/BroeckTOR10}, models the problem:
\begin{lstlisting}
0.7::buys_from_viral(P0,P1):- person(P0), person(P1), P0 != P1.
0.2::buys_from_marketing(P):- person(P).

market(P) :- person(P), not noMarket(P).
noMarket(P) :- person(P), not market(P).

buys(P):- market(P), buys_from_marketing(P).
buys(P):- friend(P,F), buys(F), buys_from_viral(P,F).

:|$\sim$| market(P). [20@1,P]
:|$\sim$| buys(P). [-5@1,P]
\end{lstlisting}
where $\mathit{person}/1$ facts describe people in the network and $\mathit{friend}/2$ facts denote friendship relationship.
A person $P$ can be marketed ($\mathit{market(P)}$) or not ($\mathit{nMarket(P)}$).
The rule with $0.7::\mathit{buys\_from\_viral}(P0,P1)$ in the head represents, in a compact way, a set of ground \pfs{} of the form $0.7::\mathit{buys\_from\_viral}(P0,P1)$ \pf{} for each pair of people $P1$ and $P2$.
These facts state that the shopping behavior of $P1$ is influenced with probability 0.7 by the behavior of $P2$.
Similarly for the probability of buying from direct marketing ($0.2::\mathit{buys\_from\_marketing}/1$).
Then, a person $P$ buys ($\mathit{buys(P)}$) a product if it is marketed and buys from the marketing or if it has a friend who buys the product and influences her.
Two \wcs{}, with the same priority level, represent the cost of two actions: marketing a particular person has a certain cost, but if someone buys the product, this gives a reward (i.e., a negative cost).
We are interested in computing, for example, the probability that a particular person or group buys the product, or that the number of people that buys the product is above a certain threshold.
\added{Overall, with optPASP we can now reason about the viral marketing scenario in a more fine-grained way, by considering different optimization criteria for the underlying ASP. }

\section{Computational Complexity Study}
\label{sec:computational_complexity}

In this section, we study the complexity of the main reasoning tasks in optimal Probabilistic Answer Set Programs.
In particular, we focus on propositional programs and consider the following decision problems.

\begin{definition}[Inference]
\label{def:decision_task_inference}
Given an optimal PASP $\optPASP$, a ground literal $q$, and a rational $\gamma$, the 
\begin{itemize}
    \item \textit{brave optimal inference} task asks whether $\upperprob^*(q) > \gamma$;
    \item \textit{cautious optimal inference} task asks whether $\lowerprob^*(q) > \gamma$.
\end{itemize}
\end{definition}

\begin{definition}[MPE]
\label{def:decision_task_mpe}
Given an optimal PASP $\optPASP$, a ground literal $e$, and a rational $\gamma$, the 
\begin{itemize}
    \item \textit{brave optimal MPE} task asks whether $\prob(\uppermpe^*(e)) \geq \gamma$;
    \item \textit{cautious optimal MPE} task asks whether $\prob(\lowermpe^*(e)) \geq \gamma$.
\end{itemize}
\end{definition}

Computational complexity results for PASP without \wcs{} (i.e., plain PASP) are well known.
In particular,~\cite{cozman2020complexity} showed that the complexity of the cautious MPE problem for HCF programs (the fragment we consider in this paper) is $\stp$, while the complexity of cautious inference is $\PP^\NP$.

\begin{observation}
Note that brave inference can be cast as cautious inference using the relation $\upperprob^*(q) = 1 - \lowerprob^*(not \ q)$.
For this reason, when we use the term inference while discussing complexity results, we refer to either brave or cautious inference.
However, this relation does not hold for $\uppermpe^*$ and $\lowermpe^*$.
Therefore the two tasks must be studied separately.
\end{observation}

We first focus on the cautious and brave MPE problems, by showing that the usage of \wcs{}
$i)$ causes a jump in complexity in the brave case (i.e., $\uppermpe(e)$) w.r.t. plain PASP while 
$ii)$ the inherited complexity of \wcs{} is absorbed in the cautious case (i.e., $\lowermpe(e)$).
So, surprisingly, the \textit{two cases have the same computational complexity for optimal PASP}.

Lastly, we focus on the brave inference task (as the cautious case can be converted to the brave one) and show that the complexity coincides with the complexity of brave inference in plain PASP.
That is, \textit{\wcs{} do not change the complexity of the inference problem}. 
To facilitate the comparison, Table~\ref{tab:complexity_results} summarizes the completeness results for both plain and optimal PASP.

\begin{table}[t]
    \renewcommand{\arraystretch}{1.2}
    \centering
    \caption{Complexity results for plain and optimal PASP. Results with a reference are proven in this paper while the remaining are taken from~\citep{cozman2020complexity}.}
    \label{tab:complexity_results}
    \begin{tabular*}{\linewidth}{@{\extracolsep{\fill}} l c c c @{}}
        \topline
        & \multicolumn{2}{c}{MPE / MPE$^*$ Complexity} & Inference Complexity \\
        \cmidrule(lr){2-3} \cmidrule(lr){4-4}
        & $\uppermpe(e)$ & $\lowermpe(e)$ & $\upperprob(q)$ / $\upperprob^*(q)$ \\
        \midrule
        Plain
        & $\NP$ (Theorem~\ref{th:upper_mpe_plain})
        & $\stp$ 
        & $\PP^\NP$ \\
        
        Weak Constraint
        & $\stp$ (Theorem~\ref{th:complexity_mpe})
        & $\stp$ (Theorem~\ref{th:complexity_mpe})
        & $\PP^\NP$ (Theorem~\ref{th:complexity_inference}) \\
        \bottomrule
    \end{tabular*}
\end{table}

\subsection{Computational Complexity of MPE}
\label{subsec:mpe_complexity}
\cite{cozman2020complexity} proved that the complexity of cautious MPE for plain PASP is complete for $\stp$.
However, to the best of our knowledge, no completeness results are available for the brave MPE problem.
To this end, the following theorem provides a completeness results for such a problem.
\added{Note that in the following proofs we assume that the logarithm can be efficiently computed and rounded in such a way that distinct probability values are mapped to distinct integers.}
\begin{theorem}
\label{th:upper_mpe_plain}
    Let $\PASP$ be a plain PASP, $e$ be a ground literal, and $\gamma \in [0,1]$ be a rational number.
    Deciding whether $\prob(\uppermpe(e))\geq\gamma$ is $\NP$-complete. 
\end{theorem}
\begin{proof}
    (Hardness) Let $\phi$ be a boolean formula in 3-CNF over a set of propositional variables $V$ of the form $C_1 \wedge \ldots \wedge C_n$, where each clause is of the form $C_i = l_i^1 \vee l_i^2 \vee l_i^3$ and each literal $l_i^j$ (with $1 \leq j \leq 3$) is either a variable $a \in V$ or its negation $\neg a$.

    Verifying the satisfiability of $\phi$ is $\NP$-complete.
    To prove hardness for brave MPE in plain PASP, we encode $\phi$ into a plain PASP $\PASP = (\ASP,\paspfacts)$, such that $\prob(\uppermpe(sat))\geq 0.5^{\vert\paspfacts\vert}$ iff $\phi$ is satisfiable.

    More precisely, $\paspfacts$ is composed of \pfs{} of the form $0.5::v$ for each variable $v \in V$, whereas $\ASP$ is of the form:
    $$
    \begin{array}{llr}
        r_1:    & asgn(v,t)        \impl v.                     & \forall v \in V \\
        r_2:    & asgn(v,f)        \impl not\ v.                & \forall v \in V \\
        r_3:    & sat(c)           \impl asgn(l,f).              & \forall c \in \phi, \neg l \in c\\
        r_4:    & sat(c)           \impl asgn(l,t).              & \forall c \in \phi, l \in c\\
        r_5:    & unsat            \impl clause(c), \naf sat(c). & \forall c \in \phi\\                       
        r_6:    & sat              \impl \naf unsat.             & \\                       
    \end{array}
    $$
    Let $\sigma$ be a possible selection, then $\sigma$ encodes a truth assignment $\tau_\sigma$ for the variables in $V$.
    More precisely, for each $v \in V$, if $v \in \sigma$, then $\tau_\sigma(v) = \top$ (i.e., $v$ is true in $\tau_\sigma$); otherwise $\tau_\sigma(v) = \bot$ (i.e., $v$ is false in $\tau_\sigma$). 
    Rules in $\ASP$ check whether a truth assignment satisfies $\phi$ or not: the atom $sat$ is derived iff such an assignment satisfies all the clauses of $\phi$; whereas the atom $\mathit{unsat}$ is derived if there is at least one clause of $\phi$ which is not satisfied.

    Thus, the world $w_\sigma$, which enforces the truth assignment $\tau_\sigma$ into $\ASP$, admits an answer set containing $sat$ iff $\tau_\sigma$ satisfies $\phi$, and so $\phi$ is satisfiable.

    Since all possible selections have the same probability, which is $0.5^{\vert\paspfacts\vert}$, and they are in one-to-one correspondence with possible truth assignments for variables in $V$, then $\prob(\uppermpe(sat))\geq 0.5^{\vert\paspfacts\vert}$ iff $\phi$ is satisfiable.

    (Membership) We prove membership to the $\NP$ class by means of an head-cycle-free ASP encoding that is coherent iff $\prob(\uppermpe(e))\geq\gamma$.
    Recall that verifying coherence of head-cycle-free ASP is $\NP$-complete~\citep{DBLP:journals/csur/DantsinEGV01}.

    Let $\PASP = (\ASP,\paspfacts)$ be a PASP and $e$ be a ground literal.
    We construct the following $\ASP_1$:
    $$
    \begin{array}{lr}
        a;na                .             & \forall p::a \in \paspfacts\\
        prob(log(p),id_a)   \impl a.      & \forall p::a \in \paspfacts\\
        prob(log(1-p),id_a) \impl not\ a. & \forall p::a \in \paspfacts\\
        H_r                 \impl B_r.    & \forall r \in \ASP\\
                            \impl \#sum\{S\}<log(\gamma). &\\
                            \impl \overline{e}.                                &
    \end{array}
    $$
    where each $id_a$ is a unique identifier for the \pf{} $p::a\in\paspfacts$, each atom $na$ is a fresh atom not appearing anywhere else, and $S$ is a ground set where for each \pf{} $p::a \in \paspfacts$ there are two elements of the form $log(p),id_a:prob(log(p),id_a)$ and $log(1-p),id_a:prob(log(1-p),id_a)$.
    \added{Note that $\overline{e}$ is the complement of $e$.}
    
    Intuitively, rules of the form $a;na$, for $p::a \in \paspfacts$, guess a possible selection $\sigma\subseteq\paspfacts^a$ and the log-probability of $\sigma$ is encoded by means of atoms over the predicate $prob/2$ deriving $prob(log(p),id_a)$ for each $a \in \sigma$ and $prob(log(1-p),id_a)$ for each $a \in \paspfacts^a \setminus \sigma$.
    Then, rules in $\ASP\subseteq\ASP_1$ guess a model of the world $w_\sigma$ and the last two constraints impose that (i) the sum of log-probability of \pfs{} cannot be less than $log(\gamma)$ and (ii) evidence $e$ cannot be false.

    Thus, there exists an \as{} of $\ASP_1$ iff there exists $\sigma\subseteq\paspfacts^a$ such that $\prob(\sigma)\geq\gamma$ and there exists $M \in \AS(w_\sigma)$ such that $e \in M$, and so the thesis follows.
\end{proof}

Theorem~\ref{th:upper_mpe_plain} completes the complexity study on MPE for plain PASP, establishing that brave MPE is \textit{easier} than the cautious MPE for plain PASP. 
We now study the complexity of MPE for optimal PASP.

First, we provide hardness for brave optimal MPE via a reduction from the well-known Propositional Abduction Problem (PAP)~\citep{DBLP:journals/jacm/EiterG95}.
Recall that a PAP is defined as a tuple of the form $\mathcal{A}=\langle V,T,H,M \rangle$, where $V$ is a set of propositional variables, $T$ is a satisfiable propositional logic theory over variables in $V$, $H \subseteq V$ is a set of hypotheses, and $M \subseteq V$ is a set of manifestations.
A \textit{solution} for $\cal{A}$ is a subset of the hypotheses $S\subseteq H$ such that $T\cup S$ is satisfiable and $T \cup S \vDash M$. 
Let $sol(\cal{A})$ be the set of solutions to PAP, then deciding whether $sol(\cal{A}) \neq \emptyset$ is $\stp$-complete~\citep{DBLP:journals/jacm/EiterG95}.

\begin{theorem}[Hardness of brave optimal MPE]
\label{th:hardness_MPE_upper}
Let $\optPASP$ be an optimal PASP, $e$ be a ground literal, and $\gamma \in [0,1]$ be a rational number.
Deciding whether $\prob(\uppermpe^*(e))\geq\gamma$ is $\stp$-hard.
\end{theorem}


\begin{proof}
Let $\mathcal{A}=\langle V,T,H,M \rangle$ be a PAP.
We assume w.l.o.g. that the theory $T$ is a 3-CNF formula over the variables in $V$. 
Recall that a 3-CNF formula is a conjunction of clauses $C_1 \wedge \ldots \wedge C_n$, where each clause is of the form $C_i = l_i^1 \vee l_i^2 \vee l_i^3$ and each $l_i^j$ (with $1 \leq j \leq 3$) is either a variable $a \in V$ or its (classical) negation $\neg a$.
With a slight abuse of notation, we refer to the theory $T$ as a set of clauses, and each clause as a set of literals (i.e., $T = \{C_1,\ldots,C_n\}$, where for each $i\in \{1,\ldots,n\}, C_i = \{l_i^1,l_i^2,l_i^3\}$).
From $\mathcal{A}$ we can construct an optimal PASP $\optPASP$ such that $\prob(\uppermpe^*(e)\geq 0.5^{\vert H\vert}$ iff $sol(\mathcal{A})\neq \emptyset$.
In detail, $\optPASP = (\ASP,\paspfacts,\weakconstraints)$ where $\paspfacts=\{0.5::s(x)\mid x \in H\}$, $\weakconstraints = \{\weakarr entailed\ \ [1@1]\}$, and $\ASP$ is of the form:
\[
    \ASP = \left\{\begin{array}{lr}
            tau(x,t);tau(x,f).                             & \forall x \in V\\
            satCl(C_i)              \impl  tau(x,t).              & \forall\ x \in V, C_i \in T \mid x \in C_i\\
            satCl(C_i)              \impl  tau(x,f).              & \forall\ x \in V, C_i \in T \mid \neg x \in C_i \\
            unsatTS                 \impl  \naf satCl(C_i).       & \forall C_i \in T\\
            unsatTS                 \impl  s(x), tau(x,f).        & \forall x \in H\\
            consistent              \impl  \naf unsatTS.          & \\
                                                                 & \\
            tau'(x,t);tau'(x,f).                             & \forall x \in V\\
            satCl'(C_i)             \impl  tau'(x,t).             & \forall\ x \in V, C_i \in T \mid x \in C_i\\
            satCl'(C_i)             \impl  tau'(x,f).             & \forall\ x \in V, C_i \in T \mid \neg x \in C_i\\
            unsatTS'                \impl  \naf satCl'(C_i).      & \forall C_i \in T\\
            unsatTS'                \impl  s(x), tau'(x,f). & \forall x \in H\\
            satTS                   \impl  \naf unsatTS'.         & \\
            notEntail               \impl  satTS, tau'(x,f).      & \forall x \in M\\
            entailed                \impl  \naf notEntail.        & \\
                                                                 & \\
            e                       \impl  consistent, entailed.  & \\            
                                    \weakarr    entailed.\ \ [1@1]  &\\
\end{array}\right\}
\]
\noindent We now prove that $\prob(\uppermpe^*(e)) \geq 0.5^{\vert\probfacts\vert}$ iff $sol(\cal{A})\neq \emptyset$.
Since each \pf{} in $\optPASP$ has probability $0.5$, each selection $\sigma\subseteq\probfacts^a$ has the same probability, which is $0.5^{\vert\probfacts\vert}$.
Thus, $\prob(\uppermpe^*(e) \geq 0.5^{\vert\probfacts\vert}$ iff there exists $\sigma \subseteq \probfacts^a$ such that $\exists I \in \optAS(w_\sigma)$ such that $e \in I$.

Let $\sigma \subseteq \probfacts^a$ be a possible selection and $S = \{x\mid s(x)\in \sigma\} \subseteq H$.
By construction, $w_\sigma$ guesses a pair of assignments $(\tau,\tau')$ (encoded by atoms over predicates $tau$ and $tau'$,  respectively), then:
\begin{itemize}
    \item $consistent$ is true in $I$ iff $\tau$ satisfies $T \cup S$; 
    \item $entailed$ is true in $I$ iff $\tau'$ satisfies the entailment $T \cup S \models M$ (i.e., if $\tau'$ satisfies $T\cup S$ and $\tau'$ satisfies also $M$); and
    \item $e$ is true in $I$ iff both $consistent$ and $entailed$ are true in $I$.
\end{itemize}  

Thus, given $I' \in \optAS(w_\sigma)$, then $e \in I'$ iff $consistent \in I'$ and $entailed \in I'$.
Since $consistent \in I'$ iff the guessed assignment $\tau$ satisfies $T\cup S$, then $\tau$ is a witness for the consistency of $T\cup S$.
On the other hand, if $entailed \in I'$, then $\cost(w_\sigma,I',1) = 1$.
So, this means that for each $I' \in \AS(w_\sigma)$, $entailed \in I'$, otherwise $I'$ would not be optimal.
Thus, for every guessed assignment $\tau'$ associated to a $I'\in \optAS(w_\sigma)$, $\tau'$ satisfies $T\cup S \models M$, and so $entailed \in I$ iff $T \cup S \models M$ holds.
Since $consistent \in I'$ iff $T\cup S$ is consistent and $entailed \in I'$ iff $T \cup S \models M$ holds, then, we can conclude that $e \in I$ iff $S \in sol(\cal{A})$ and so  $\prob(\uppermpe^*(e)) \geq 0.5^{\vert\probfacts\vert}$ iff $sol(\cal{A})\neq\emptyset$.
\end{proof}

\begin{theorem}[Hardness of cautious optimal MPE]
\label{th:hardness_MPE_lower}
Let $\optPASP$ be an optimal PASP, $e$ be a ground literal, and $\gamma \in [0,1]$ be a rational number. Deciding whether $\prob(\lowermpe^*(e))\geq\gamma$ is $\stp$-hard.
\end{theorem}
\begin{proof}
The proof follows directly by observing that each PASP is indeed an optimal PASP where the underlying ASP contains no \wcs{}.
Deciding whether $\prob(\lowermpe^*(e))\geq\gamma$ for a PASP is $\stp$-hard~\citep{cozman2020complexity} and so the thesis follows.
\end{proof}

From Theorem~\ref{th:hardness_MPE_upper} and~\ref{th:hardness_MPE_lower} we can observe that \wcs{} cause a jump in complexity for brave MPE, but they do not introduce further complexity for cautious MPE.
Indeed, both problems are hard for $\stp$. 

We now provide memberships, which give us the completeness for cautious/brave MPE for optimal PASP.
In both cases, we rely on \ASPQ{}~\citep{DBLP:journals/tplp/AmendolaRT19,DBLP:journals/tplp/MazzottaRT24} 
which allows modeling this problem in a very natural way. 

\begin{theorem}[Membership of brave optimal MPE]
\label{th:membership_MPE_upper}
Let $\optPASP$ be an optimal PASP, $e$ be a ground literal, and $\gamma \in [0,1]$ be a rational number.
Deciding whether $\prob(\uppermpe^*(e))\geq\gamma$ is in $\stp$.
\end{theorem}

\begin{proof}
For an optimal PASP $\optPASP = (\ASP,\probfacts,\weakconstraints)$, deciding whether $\prob(\uppermpe^*(e))\geq\gamma$ is equivalent at verifying that there exists a selection $\sigma \subseteq\probfacts^a$ such that $\prob(\sigma)\geq\gamma$ and there exists $M \in \optAS(w_\sigma)$ such that $e \in M$.

To this end, we provide an \ASPQ encoding that verifies the existence of such $\sigma$.
More precisely, we can encode this problem into an \ASPQ, namely $\Pi$, of the form $\exists^{st} \ASP_1\exists^{st} \ASP_2: C$ where $\ASP_1$ is made of the following rules for each \pf{} $p::a\in\paspfacts$: 
$$
\begin{array}{rcl}
     a;na.                      &       & \\
     prob(log(p),id_a)               & \impl & a.\\
     prob(log(1-p),id_a)    & \impl & not\ a.\\
\end{array}
$$
where $id_a$ is a unique identifier for the \pf{} $p::a$ and $na$ is a fresh atom not appearing anywhere else. Finally, $\ASP_2 = \ASP\cup \weakconstraints$, and $C$ is of the form:
$$
\begin{array}{rl}
     \impl & \overline{e}.\\
     \impl & \#sum\{S\}< log(\gamma).\\
\end{array}
$$
where $S$ is a ground set where for each \pf{} $p::a \in \paspfacts$ there are two elements of the form $log(p),id_a:prob(log(p),id_a)$ and $log(1-p),id_a:prob(log(1-p),id_a)$.
From Theorem 5 by~\cite{DBLP:journals/tplp/MazzottaRT24}, verifying the coherence of $\Pi$ is $\stp$-complete.
So, this gives us the membership to $\stp$.

By construction, the program $\ASP_1$ guesses a set $S \subseteq \paspfacts^a$ and then entails:
\begin{itemize}
    \item an atom $prob(log(p),id_a)$ iff $p::a \in \paspfacts$ and $a \in S$; and
    \item an atom $prob(log(1-p),id_a)$ iff $p::a \in \paspfacts$ and $a \notin S$.
\end{itemize} 

Thus, $S$ corresponds to a possible selection, and atoms over the predicate $prob/2$ encode the log-probability of such a selection.
So, there is a one-to-one correspondence between possible selections and the answer sets of program $\ASP_1$.   
According to the semantics of \ASPQ, $\Pi$ is coherent iff there exists $M_1 \in \optAS(\ASP_1)$ such that $\exists^{st} \ASP_2': C$ is coherent, with $\ASP_2' = \ASP_2 \cup \fix{\ASP_1}{M_1}$. 

Let $M_1 \in AS(\ASP_1) = \optAS(\ASP_1)$ and $\sigma$ the corresponding selection then, by construction, $\fix{\ASP_1}{M_1}$ is of the form:
$$
\begin{array}{lr}
    a.                     & \forall p::a \in \paspfacts\ s.t.\ a \in M_1      \\
    na.                    & \forall p::a \in \paspfacts\ s.t.\ a \notin M_1      \\
    prob(log(p),id_a).     & \forall p::a \in \paspfacts\ s.t.\ a \in M_1      \\
    prob(log(1-p),id_a).   & \forall p::a \in \paspfacts\ s.t.\ a \notin M_1   \\
                                    &                                               \\
    \impl a.                    & \forall p::a \in \paspfacts\ s.t.\ a \notin M_1   \\
    \impl na.                    & \forall p::a \in \paspfacts\ s.t.\ a \in M_1   \\
    \impl prob(log(p),id_a).    & \forall p::a \in \paspfacts\ s.t.\ a \notin M_1   \\
    \impl prob(log(1-p),id_a).  & \forall p::a \in \paspfacts\ s.t.\ a \in M_1   \\
    
\end{array}
$$
We can observe that, by construction, fresh atoms of the form $na$, for some $p::a \in \paspfacts$, \pfs{} and atoms over the predicate $prob/2$ do not appear in the head of any rule of $\pasprules \subseteq \ASP_2'$. So, this means that atoms appearing in the constraints of $\fix{\ASP_1}{M_1}$ are false in every answer set of $\ASP_2'$ and so such constraints are trivially satisfied. Thus, we can simplify $\ASP_2'$ to $\ASP\cup \weakconstraints \cup Facts$, where $Facts$ is: 

$$
\begin{array}{lr}
    a.                      & \forall p::a \in \paspfacts\ s.t.\ a \in M_1 \\
    na.                     & \forall p::a \in \paspfacts\ s.t.\ a \notin M_1 \\
    prob(log(p),id_a).      & \forall p::a \in \paspfacts\ s.t.\ a \in M_1 \\
    prob(log(1-p),id_a).    & \forall p::a \in \paspfacts\ s.t.\ a \notin M_1 \\
\end{array}
$$

Moreover, we observe that fresh atoms of the form $na$, for some $p::a \in \paspfacts$, and the atoms over the predicate $prob/2$ do not appear in anywhere else in $\ASP_2'$ and so they do not affect the answer sets of the program $\ASP_2'$. 

Let $A$ be the following set of atoms:
$$
A=\left\{\begin{array}{lr}
    na.                   & \forall p::a \in \paspfacts\ s.t.\ a \notin M_1 \\
    prob(log(p),id_a)     & \forall p::a \in \optPASP\ s.t.\ a \in M_1      \\
    prob(log(1-p),id_a)   & \forall p::a \in \optPASP\ s.t.\ a \notin M_1   \\
\end{array}
\right.
$$
Then, we can remove facts over atoms in $A$ from $\ASP_2'$, obtaining the world $w_\sigma$, and so answer sets of $\ASP_2'$ will be of the form $A \cup M_2$ where $M_2 \in AS(w_\sigma)$. 
Since atoms in $A$ do not appear in any \wc{} of $\weakconstraints$, then $A \cup M_2 \in \optAS(\ASP_2')$ iff $M_2 \in \optAS(w_\sigma)$.

According to the coherence of \ASPQ, $\exists^{st} \ASP_2': C$ is coherent iff there exists $M_2' \in \optAS(\ASP_2')$ such that $C'=C \cup \fix{\ASP_2'}{M_2'}$ is coherent.

Let $M_2' \in \optAS(\ASP_2')$, then $M_2' = A \cup M_2$ and $M_2 \in \optAS(w_\sigma)$.
By construction, the program $C'$ is coherent iff the constraints in $C$ are satisfied w.r.t. $M_2'$.
Thus, $C'$ is coherent iff $M_2\models e$ and the sum of the log-probability of \pfs{} (encoded by atoms over the predicate $prob/2$ in $A$) is greater than or equal to $log(\gamma)$. 

Thus, $\Pi$ is coherent iff there exists $\sigma\subseteq \paspfacts^a$ such that $\prob(\sigma)\geq\gamma$ and there exists $M \in \optAS(w_\sigma)$ such that $M\models e$, that is $\prob(\uppermpe^*(e))\geq\gamma$. Thus, the thesis follows.
\end{proof}

Based on the approach of the proof of Theorem~\ref{th:membership_MPE_upper} we can also provide a membership result for cautious MPE.  
\begin{theorem}[Membership of cautious optimal MPE]
\label{th:membership_MPE_lower}
Let $\optPASP$ be an optimal PASP, $e$ be a ground literal, and $\gamma \in [0,1]$ be a rational number.
Deciding whether $\prob(\lowermpe^*(e))\geq\gamma$ is in $\stp$.
\end{theorem}


\begin{proof}
For an optimal PASP $\optPASP = (\ASP,\probfacts,\weakconstraints)$, deciding whether $\prob(\lowermpe^*(e))\geq\gamma$ is equivalent to verifying that there exists a selection $\sigma \subseteq\paspfacts$ such that $\prob(\sigma)\geq\gamma$ and for each $M \in \optAS(w_\sigma)$, $e \in M$.
Thus, similarly to Theorem~\ref{th:hardness_MPE_upper}, we provide an \ASPQ encoding that verifies the existence of such $\sigma$.
More precisely, we can encode such a problem into an \ASPQ $\Pi$ of the form $\exists^{st} \ASP_1\exists^{st} \ASP_2: C$ where $\ASP_1$ is equal to $\ASP_1$ of~\ref{th:membership_MPE_upper}.
Finally, $\ASP_2 = \ASP \cup \weakconstraints \cup \{\weakarr e\ [1@lmin-1]\}$, with $lmin$ being the lowest level among \wcs{} in $\weakconstraints$, and $C$ is 
as in Theorem~\ref{th:membership_MPE_upper}.
As in the previous case, verifying the coherence of $\Pi$ is $\stp$-complete~\citep{DBLP:journals/tplp/MazzottaRT24}. 

By construction, the program $\ASP_1$ guesses a set $S \subseteq \paspfacts$ and then entails:
\begin{itemize}
    \item an atom $prob(log(p),id_a)$ iff $p::a \in \optPASP$ and $a \in S$; and
    \item an atom $prob(log(1-p),id_a)$ iff $p::a \in \optPASP$ and $a \notin S$.
\end{itemize} 

Thus, $S$ corresponds to a possible selection, and the atoms over the predicate $prob/2$ encode the log-probability of such a selection.
So, there is a one-to-one correspondence between the possible selections and the answer sets of the program $\ASP_1$.   
According to the semantics of \ASPQ, $\Pi$ is coherent iff there exists $M_1 \in \optAS(\ASP_1)$ such that $\exists^{st} \ASP_2': C$ is coherent, with $\ASP_2' = \ASP_2 \cup \fix{\ASP_1}{M_1}$. 

Let $M_1 \in AS(\ASP_1) = \optAS(\ASP_1)$, and $\sigma$ be the corresponding selection.
By construction $\fix{\ASP_1}{M_1}$ is of the form:
$$
\begin{array}{lr}
    a.                          & \forall p::a \in \paspfacts\ s.t.\ a \in M_1      \\
    na.                         & \forall p::a \in \paspfacts\ s.t.\ a \notin M_1      \\
    prob(log(p),id_a).          & \forall p::a \in \paspfacts\ s.t.\ a \in M_1      \\
    prob(log(1-p),id_a).        & \forall p::a \in \paspfacts\ s.t.\ a \notin M_1   \\
                                &                                               \\
    \impl a.                    & \forall p::a \in \paspfacts\ s.t.\ a \notin M_1   \\
    \impl na.                   & \forall p::a \in \paspfacts\ s.t.\ a \in M_1   \\
    \impl prob(log(p),id_a).    & \forall p::a \in \paspfacts\ s.t.\ a \notin M_1   \\
    \impl prob(log(1-p),id_a).  & \forall p::a \in \paspfacts\ s.t.\ a \in M_1   \\
    
\end{array}
$$
As in the previous case, we can observe that, by construction, atoms of the form $na$, for some $p::a \in \paspfacts$, \pfs, and atoms over the predicate $prob/2$ do not appear in the head of any rule in $\ASP \subseteq \ASP_2'$.
So, this means that atoms appearing in the constraints of $\fix{\ASP_1}{M_1}$ are false in every answer set of $\ASP_2'$ and so such constraints are trivially satisfied.
Thus, we can simplify $\ASP_2'$ to $\ASP \cup \weakconstraints \cup \{\weakarr e\ [1@lmin-1]\} \cup \mathit{Facts}$, where $\mathit{Facts}$ is of the form:
$$
\begin{array}{lr}
    a.                      & \forall p::a \in \paspfacts\ s.t.\ a \in M      \\
    na.                     & \forall p::a \in \paspfacts\ s.t.\ a \notin M      \\
    prob(log(p),id_a).      & \forall p::a \in \paspfacts\ s.t.\ a \in M      \\
    prob(log(1-p),id_a).    & \forall p::a \in \paspfacts\ s.t.\ a \notin M   \\
\end{array}
$$

Moreover, we observe that atoms of the form $na$, for some $p::a \in \paspfacts$, and atoms over the predicate $prob/2$ do not appear in anywhere else in $\ASP_2'$ and so they do not affect the answer sets of the program $\ASP_2'$. 

Let $A$ be the following set of atoms:
$$
A=\left\{\begin{array}{lr}
    na                  & \forall p::a \in \optPASP\ s.t.\ a \in M      \\
    prob(log(p),id_a)   & \forall p::a \in \optPASP\ s.t.\ a \in M      \\
    prob(log(1-p),id_a) & \forall p::a \in \optPASP\ s.t.\ a \notin M   \\
\end{array}
\right..
$$
Then, we can remove from $\ASP_2'$ facts over the atoms in $A$, obtaining the program $P = w_\sigma \cup \{\weakarr e\ [1@lmin-1]\}$
So, each \as{} of $\ASP_2'$ is of the form $A \cup M$ where $M \in AS(P) = AS(w_\sigma)$, since $P$ extends the world $w_\sigma$ with a \wc{}.

According to the semantics of \ASPQ{}, $\exists^{st} \ASP_2': C$ is coherent iff there exists $M_2 \in \optAS(\ASP_2')$ such that $C\cup \fix{\ASP_2'}{M_2}$ is coherent. 
The constraints in $C$ impose that the sum of the log-probabilities of \pfs{} (encoded by atoms over the predicate $prob/2$ in $A$) must be greater than or equal to $log(\gamma)$ and $e$ must be true w.r.t. $M_2$.

Thus, we now show that there exists such $M_2$ iff the evidence $e$ is true in every optimal answer set of $w_\sigma$, which means $\prob(\lowermpe^*(e))\geq\gamma$ and so the thesis follows.

Let $M_2 \in \optAS(\ASP_2')$, then $M_2 = A \cup M$ where $M \in \optAS(P)$. Then we show that (i) $M \in \optAS(w_\sigma)$ and (ii) $M\models e$ iff $e$ is true in every $M' \in \optAS(w_\sigma)$.

By construction we know that:
\begin{itemize}
    \item for each $l \geq lmin$, $\cost(P,M,l) = \cost(w_\sigma,M,l)$; and
    \item if $e \notin M$, then $\cost(P,M,lmin-1) = 0$, otherwise $\cost(P,M,lmin-1) = 1$.
\end{itemize}

So, if $M \notin \optAS(w_\sigma)$ (i.e., $M$ is an optimal answer set of $P$ but it is not optimal for $w_\sigma$), then $M$ is dominated by $M'\in \optAS(w_\sigma) \subseteq AS(w_\sigma) = \AS(P)$, and so $M$ would not be optimal for $P$, which is a contradiction.
So $M \in \optAS(w_\sigma)$. 

Now, let us assume that there exists $M' \in \optAS(w_\sigma)$ such that $M'\not\models e$ and the evidence $e$ is true in $M$ (i.e., $M\models e$).
Since $\AS(P) = \AS(w_\sigma)$, then $M' \in \AS(P)$ and $\cost(P,M',lmin-1) = 0 < 1 = \cost(P,M,lmin-1)$.
Thus, $M$ is dominated by $M'$ since $M \in \optAS(w_\sigma)$, and so, for each $l \geq lmin$, $\cost(P,M',l) = \cost(P,M,l)$ and $M \not\in \optAS(P)$.
This is a contradiction, since $M$ is an optimal answer set of $P$.

Thus, we can conclude that $e \in M$ iff for each $M' \in \optAS(w_\sigma)$, $e \in M'$, and the thesis follows.
\end{proof}

\begin{theorem}[Complexity of cautious and brave optimal MPE]
\label{th:complexity_mpe}
    Given an optimal PASP $\optPASP$, a ground literal $e$, and a rational number $\gamma \in [0,1]$, deciding whether $(i)$ $\prob(\lowermpe^*(e)) \geq \gamma$ is $\stp$-complete and $(ii)$ $\prob(\uppermpe^*(e)) \geq \gamma$ is $\stp$-complete.
\end{theorem}
\begin{proof}
    The proof follows from theorems~\ref{th:hardness_MPE_upper}-\ref{th:membership_MPE_lower}.
\end{proof}

\subsection{Computational Complexity of Inference}
\label{subsec:inference_complexity}
We now study the complexity of the brave inference problem (Definition~\ref{def:decision_task_inference}).
We start from the hardness result and then also provide a membership result, thus proving completeness.
Recall that brave inference can be converted to cautious inference and vice versa, so studying one is enough.
\added{This is because $\upperprob(q) = 1 - \lowerprob(not \ q)$ for a query atom $q$.
Also note that for MPE, converting the lower state into the upper state is not possible in general, as the latter is a world in which the query is cautiously or bravely entailed.}

\begin{theorem}[Hardness of brave optimal inference]
\label{th:upperprob_hardness}
    Let $\optPASP$ be an optimal PASP, $q$ be a ground literal, and $\gamma$ be a rational number in $[0,1]$.
    Verifying whether $\upperprob^*(q) > \gamma$ is hard for $\PP^\NP$.
\end{theorem}
\begin{proof}
Hardness directly follows from the inference problem in PASP.
Indeed, every PASP is an optimal PASP where the set of \wcs{} is empty, so the complexity of the inference problem is at least difficult as in PASP, that is, $\PP^\NP$~\citep{cozman2020complexity}. 
\end{proof}

Before considering membership of the brave inference problem for optimal PASP, we first recall the complexity of brave reasoning in HCF programs with \wcs{} (i.e., the ASP fragment considered in this paper), which is instrumental in what follows.
\begin{theorem}[\cite{DBLP:journals/tkde/BuccafurriLR00}]
\label{thm:brave_asp_weak}
Let $\ASP$ be an HCF program with \wcs{} and $q$ be ground literal.
Verifying whether there exists $M \in \optAS(\ASP)$ such that $M\models q$ is $\dtp$-complete.    
\end{theorem}

\begin{theorem}[Membership of brave optimal inference]
\label{th:upperprob_membership}
    Let $\optPASP$ be an optimal PASP, $q$ be an atom, and $\gamma \in [0,1]$ be a rational number.
    Then, verifying whether $\upperprob^*(q) > \gamma$ is in $\PP^\NP$.
\end{theorem}

\begin{proof}
To prove this, we can apply, with slight modifications, the methodology employed by~\cite{DBLP:journals/jair/CozmanM17} to prove the complexity of brave inference for plain PASP.

Let $\optPASP = (\ASP,\paspfacts,\weakconstraints)$ be an optimal PASP.
We construct a non-deterministic Turing Machine $TM_{\optPASP}$ such that, for each \pf{} $p_i::a_i \in \paspfacts$, with $p_i = \mu_i/\nu_i$ for some (smallest) integers $\mu_i$ and $\nu_i$, it chooses one among $\nu_i$ computation paths. 
For $\mu_i$ of these computation paths, $a_i$ is considered true, while for the remaining $\nu_i-\mu_i$, $a_i$ is considered false. 
This means that such a machine has $\prod_i \nu_i$ computation paths, and for each selection $\sigma \subseteq \paspfacts^a$, there are $\prod_{a_i\in\sigma} \mu_i\ \cdot \prod_{a_i \in \paspfacts^a\setminus \sigma} (\nu_i-\mu_i)$ computation paths. 
Each of these paths is polynomial w.r.t. the input size and returns yes iff $\exists M \in \optAS(w_\sigma)$ such that $M\models q$.

Let $\mathcal{S} = \{\sigma \subseteq \paspfacts^a \mid \exists M \in \optAS(w_\sigma) \text{ s.t. } M\models q\}$, i.e., the set of selections $\sigma$ such that the query $q$ is true in at least one optimal \as{} of the world $w_\sigma$ identified by $\sigma$.
Then, the number of accepting paths of $TM_{\optPASP}$ is equal to 
$$
\sum\limits_{\sigma\in\mathcal{S}} (\prod_{a_i\in\sigma} \mu_i\ \cdot \prod_{a_i \in \paspfacts^a\setminus \sigma} (\nu_i-\mu_i)).
$$
If we divide this quantity by the total number of paths, we obtain the probability of the query $q$.
That is,
$$
\upperprob(q) = \frac{\sum\limits_{\sigma\in\mathcal{S}} (\prod_{a_i\in\sigma} \mu_i\ \cdot \prod_{a_i \in \paspfacts^a \setminus \sigma} (\nu_i-\mu_i))}{\prod_{a_i \in \paspfacts^a} \nu_i}.
$$

This means that $\upperprob(q) > \gamma$ if and only if the number of accepting paths of $TM_{\optPASP}$ is greater than $\gamma \cdot \prod_{a_i \in \paspfacts^a} \nu_i$.

If $\gamma = 1/2$, then $\upperprob(q)>\gamma$ if and only if more than half of the computational paths are accepting.
For a generated selection, verifying the existence of an optimal \as{} is $\dtp$ (Theorem~\ref{thm:brave_asp_weak}).
Thus, for $\gamma = 1/2$ the problem is in $\PP^{\dtp}$.


We now reduce the general case, i.e., with an arbitrary value of $\gamma$, to the specific case $\gamma = 1/2$, obtain membership in $\PP^{\dtp}$.
\added{First, consider $\gamma > 1/2$. Then}
construct an optimal PASP $\optPASP' = (\ASP',\paspfacts',\weakconstraints)$, where
$\ASP' = \ASP \cup \{\mathit{aux}_2\impl q,\mathit{aux}_1\}$, and
$\paspfacts' = \paspfacts \cup \{1/(2\gamma)::\mathit{aux}_1\}$,
with fresh atoms $\mathit{aux}_1$ and $\mathit{aux}_2$ not appearing elsewhere,
and prove that $\upperprob(\mathit{aux}_2) > 1/2$ if and only if $\upperprob(q) > \gamma$.

Selections for $\optPASP'$ are obtained from selections for $\optPASP$ by including or not $\mathit{aux}_1$.
Therefore, for each selection $\sigma \subseteq \paspfacts^a$ for $\optPASP$, we have two possible selections,
$\sigma_1 = \sigma$, where $\mathit{aux}_1$ is not included, and $\sigma_2 = \sigma \cup \{\mathit{aux}_1\}$, where $\mathit{aux}_1$ is included.
The corresponding worlds $w_{\sigma_1}$ and $w_{\sigma_2}$ preserve all the optimal answer sets of $w_\sigma$ as $\mathit{aux}_1$ and $\mathit{aux}_2$ appear only in the new rule $\mathit{aux}_2\impl q,\mathit{aux}_1$ and they do not appear in any \wcs{}.
For each $M \in \optAS(w_\sigma)$ we have: 
1) $M \in \optAS(w_{\sigma_1})$ since $\mathit{aux}_1 \notin \sigma_1$ and so $\mathit{aux}_2\notin M$; and
2) $M \cup \{\mathit{aux}_1,\mathit{aux}_2\} \in \optAS(w_{\sigma_2})$ if and only if $M\models q$.
Thus,
\begin{equation*}
\begin{split}
\upperprob(\mathit{aux}_2) &= \sum\limits_{\sigma\subseteq\paspfacts^a\mid \exists M \in \optAS(w_{\sigma}), M\models q} \prob(\sigma_2)\\
&= \sum\limits_{\sigma\subseteq\paspfacts^a\mid \exists M \in \optAS(w_{\sigma}), M\models q} \frac{1}{(2\gamma)}\cdot\prob(\sigma)
\\
&= \frac{1}{(2\gamma)} \cdot \sum\limits_{\sigma\subseteq\paspfacts^a\mid \exists M \in \optAS(w_\sigma), M\models q} \prob(\sigma) \\
&= \frac{1}{(2\gamma)} \cdot \upperprob(q).
\end{split}
\end{equation*}


Thus, $\upperprob(\mathit{aux}_2)> 1/2$ if and only if $1/(2\gamma)\cdot \upperprob(q)> 1/2$ i.e., $\upperprob(q)>\gamma$.
\added{If $\gamma < 1/2$, then set the probability of $aux1$ to $\frac{1-2\gamma}{2 - 2\gamma}$ and proceed as before.} 
Thus, deciding whether $\upperprob(q)>\gamma$ is in $\PP^{\dtp}$.
As observed by~\cite{cozman2020complexity}, $\PP^{\dtp} = \PP^{\NP}$ and so the thesis follows.
\end{proof}

\begin{theorem}\label{th:complexity_inference}
    Let $\optPASP$ be an optimal PASP, $q$ be a ground literal, and $\gamma \in [0,1]$ be a rational number.
    Then, verifying whether $\upperprob^*(q) > \gamma$ is $\PP^{\NP}$-complete.
\end{theorem}
\begin{proof}
    Follows from Theorem~\ref{th:upperprob_hardness} and~\ref{th:upperprob_membership}.
\end{proof}

\section{Implementation of Inference and MPE in Optimal PASP}
\label{sec:implementation}

We now show how inference and the computation of the MPE in optimal PASP can be naturally encoded as a 2AMC taks.
After this, we consider an \ASPQ-based implementation for the MPE task.

\subsection{Encoding Inference as a 2AMC Task}
\label{subsec:inference_encoding}

Here we consider an optimal PASP with 
query $q$, 
\pfs{} $\probfacts$,
and \wcs{} $\weakconstraints$,
where, w.l.o.g., each \wc{} is of the form $:\sim a. [w,id_a]$, $a$ is a ground atom, and $w > 0$ (the weight).
We omit the level $l$, as multiple levels of \wcs{} can be rearranged into only one level and negative weights can be rewritten as positive, see~\citep{DBLP:journals/tplp/AlvianoD16,DBLP:journals/tkde/BuccafurriLR00}.
Furthermore, each \wc{} of the form $:\sim l_1,\ldots,l_n \quad [w,t]$ can be rewritten as a normal rule of the form $v(w,t)\impl l_1,\ldots,l_n$ and a \wc{} of the form $:\sim v(w,t) \quad [w,t]$.
This will be instrumental in what follows, as our implementation assumes that weights are associated with atoms.

To perform inference, the two layers of AMC (Definition~\ref{def:2amc}) are instantiated as follows.

\paragraph{Inner Layer.}
For the inner layer $A_i$ we have:
\begin{itemize}
  \item as semiring, $\mathcal{R}^{i} = (\mathbb{N}^3, \oplus, \otimes, (\infty,0,0), (0,1,1))$;
  \item as variables $\base \setminus \probfacts$
  \item as weight function: 
    \begin{equation*}
        w_i(a) =
        \begin{dcases*}
          (0,0,1) & if $a = not \ q$ \\
          (w(a),1,1) & if $a \in W$ \\
          (0,1,1) & otherwise
        \end{dcases*}
    \end{equation*}
\end{itemize}
where $(a_0,a_1,a_2) \otimes (b_0,b_1,b_2) = (a_0+b_0, a_1 \cdot b_1, a_2 \cdot b_2)$ and $w(a)$ is the weight (cost) associated with atom $a$,
and
\begin{equation*}
    (a_0,a_1,a_2) \oplus (b_0,b_1,b_2) =
    \begin{dcases*}
      (a_0,a_1,a_2) & if $a_0 < b_0$ \\
      (b_0,b_1,b_2) & if $b_0 < a_0$ \\
      (a_0,a_1 + b_1, a_2 + b_2) & if $a_0 = b_0$.
    \end{dcases*}
\end{equation*}
The first component stores the weights, the second counts the number of models where the query $q$ is true, while the last counts all the models.
As the transformation function, we use:
\begin{equation*}
    f_{io}((n_0,n_1,n_2)) =
    \begin{dcases*}
    (0,0) & \text{if } $n_1 = 0$ \\
    (0,1) & \text{if } $n_2 > n_1 > 0$ \\
    (1,1) & \text{if } $n_1 = n_2 > 0$
    \end{dcases*}
\end{equation*}
Remember that $n_1$ cannot be larger than $n_2$.

\paragraph{Outer Layer.}
For the outer layer $A_o$ we have:
\begin{itemize}
  \item as semiring, $\mathcal{R}^{o} = ([0, 1]^2, +, \cdot,(0, 0),(1, 1))$, i.e., a double probability semiring extended to two dimensions;
  \item as variables $\probfacts$;
  \item as weight function: 
    \begin{equation*}
        w_m(a) =
        \begin{dcases*}
          (p,p) & if $a = f$ with p.f. $p::f$, \\
          (1-p,1-p) & if $a = not \ f$ with p.f. $p::f$, \\
          (1,1) & otherwise.
        \end{dcases*}
    \end{equation*}
    where p.f. stands for \pf{}.
\end{itemize}

\begin{theorem}
\label{th:commutative_inner_layer}
$\mathcal{R}^{i}$ is commutative. 
\end{theorem}
\begin{proof}
To prove that \(\mathcal{R}^{i}\) is commutative, we have to show that both $\oplus$ and $\otimes$ are commutative.
Let \(a = (a_0, a_1, a_2)\) and \(b = (b_0, b_1, b_2)\) be arbitrary elements in \(\mathbb{N}^3\).
For $\oplus$, we have to show that \(a \oplus b = b \oplus a\).
We have three cases:
1) $(a_0 < b_0)$, 2) $(b_0 < a_0)$, and 3) $(a_0 = b_0)$.
Cases 1) and 2) follow from the definition.
For case 3) $a \oplus b = (a_0, a_1 + b_1, a_2 + b_2)$, $b \oplus a = (b_0, b_1 + a_1, b_2 + a_2)$, addition in $\mathbb{N}$ is commutative, and $a_0 = b_0$, so the resulting tuples are identical.

For $\otimes$, we have to show, as before, that $a \otimes b = b \otimes a$.
For $a \otimes b = (a_0 + b_0, a_1 \cdot b_1, a_2 \cdot b_2)$ and $b \otimes a = (b_0 + a_0, b_1 \cdot a_1, b_2 \cdot a_2)$.
Addition and multiplication on $\mathbb{N}$ are commutative,
thus $a \otimes b = b \otimes a$.    
\end{proof}

\begin{theorem}
\label{th:commutative_outer_layer}
$\mathcal{R}^{o}$ is commutative.
\end{theorem}
\begin{proof}
This is trivial, as it is a duplication of the probability semiring~\citep{10.1016/j.jal.2016.11.031}, which is commutative.
\end{proof}

We now prove that the transformation function $f_{io}$ is a monoid homomorphism between the inner and outer semiring.
This is crucial to ensure an efficient implementation (see ~\citep{DBLP:journals/tplp/KieselTK22}).

\begin{definition}[Monoid Homomorphism]
\label{def:monoid_homomorphism}
Given two monoids
$\mathcal{M}_0 = (D_0, \otimes_0, e_0)$
$\mathcal{M}_1 = (D_1, \otimes_1, e_1)$ and
where, for $i \in \{0,1\}$, $D_i$ are the domains, $\otimes_i$ are binary operations with neural elements $e_i$,
a function $f : \mathcal{M}_0 \to \mathcal{M}_1$ is a monoid homomorphism if $\forall m_a,m_b \in \mathcal{D}_0$,
1) $f(e_0) = e_1$
and
2) $f(m_a \otimes_0 m_b) = f(m_a) \otimes_1 f(m_b)$.
\end{definition}

\begin{theorem}
\label{th:monoid_homomorphism}
$f_{io}$ is a monoid homomorphism between
$\mathcal{M}^{i} = (\mathbb{N}^3_0, \otimes, (0,1,1))$ (obtained from $\mathcal{R}^{i}$)
and
$\mathcal{M}^{o} = ([0, 1]^2, \cdot, (1, 1))$ (obtained from $\mathcal{R}^{o}$)
\end{theorem}
\begin{proof}
We have to prove that
1) $f_{io}((0,1,1)) = (1, 1)$
and
2) $\forall m_1, m_2 \in \mathcal{M}^{i}, f_{io}(m_1 \otimes m_2) = f_{io}(m_1) \cdot f_{io}(m_2)$.
For 1) we are in the case of $n_1 = n_2 = 1 > 0$, thus, $f_{io}((0,1,1)) = (1, 1)$ by definition.
For 2) we have to consider three possible cases.

Case 1, where $m_1 = (a,0,0)$ and $m_2 = (b,0,0)$.
Then $m_1 \otimes m_2 = (a+b,0,0)$.
$f_{io}((a+b,0,0)) = (0,0)$.
In the other direction,
$f_{io}((a,0,0)) = (0,0)$,
$f_{io}((b,0,0)) = (0,0)$,
$(0,0) \cdot (0,0) = (0,0)$, and the two results match.

Case 2, where 
$m_1 = (a,v_1,v_1)$ and $m_2 = (b,v_2,v_2)$ with $v_1,v_2 > 0$.
Then $m_1 \otimes m_2 = (a+b,v_1 \cdot v_2,v_1 \cdot v_2)$, and
$f_{io}(m_1 \otimes m_2) = (1,1)$ as the last two elements are equal and $> 0$.
In the other direction,
$f_{io}(m_1) = (1,1)$ and
$f_{io}(m_2) = (1,1)$,
so $(1,1) \cdot (1,1) = (1,1)$, and the two results match. 

Case 3, where 
$m_1 = (a,v_1,v_2)$ with $0 < v_1 < v_2$ and
$m_2 = (b,v_3,v_4)$ with $0 < v_3 < v_4$.
Then $m_1 \otimes m_2 = (a+b,v_1 \cdot v_3,v_2 \cdot v_4)$.
Note that $v_1 \cdot v_3 < v_2 \cdot v_4$ so
$f_{io}(m_1 \otimes m_2) = (0,1)$.
In the other direction,
$f_{io}(m_1) = (0,1)$ and
$f_{io}(m_2) = (0,1)$,
so
$(0,1) \cdot (0,1) = (0,1)$, and the two results match. 
The other case where, for either $m_1$ or $m_2$, both the second and third elements of the tuple are equal can be similarly proven.
\end{proof}

\begin{example}
\label{ex:semiring_inference_operations_on_program}
Consider Example~\ref{ex:propositional_optpasp} and query $q$.
Here, in the innermost layer we consider the atoms $\{q,nq,qr\}$ while in the outermost $\{a,b\}$.
Assume we fix the atom $b$ in the outermost layer.
For clarity, assume $qr$ has cost 2.

We have two answer sets: 
$A_0 = \{q\}$ and $A_1 = \{nq,qr\}$.
For $A_0$, the atom $q$ is associated with the tuple $(0,1,1)$.
For $A_1$, 
the atom $nq$ is associated with the tuple $(0,1,1)$
and 
the atom $qr$ is associated with the tuple $(2,1,1)$.
Furthermore, there is an additional tuple to consider, $(0,0,1)$, as the query is false (we do not explicitly include such an atom in the answer sets).
Overall, we have
$(0,1,1) \otimes (2,1,1) \otimes (0,0,1) = (2,0,1)$.
\added{Then, $(0,1,1) \oplus (2,0,1) = (0,1,1)$ as $1 < 2$ (the first element of the tuple).}
\added{After applying the transformation function, we get $f((0,1,1)) = (1,1)$, as the second element of the tuple (which counts the number of answer sets with the query true) is 1.
In the outer layer, we have
$(0.3,0.3) \cdot (0.8,0.8) \cdot (0,0) = (1,1)$, so a contribution of 0.24 to both the lower and upper probability.}

Similarly, If we consider the atom $a$ in the outermost layer, we have only one answer set $\{a,q\}$ whose atoms are both associated with the tuple $(0,1,1)$.
Then, $(0,1,1) \otimes (0,1,1) = (0,1,1)$.
We apply the transformation function on this and get $f_{io}((0,1,1)) = (1,1)$.
In the outer layer, we have
$(0.7,0.7) \otimes (0.2,0.2) \otimes (1,1) = (0.14,0.14)$ so a contribution to both the lower and upper probability.
The considerations for the two remaining selections are analogous.
\end{example}

\subsection{Encoding MPE as a 2AMC Task}
\label{subsec:mpe_encoding}

The 2AMC encoding can be obtained with minimal modifications from the 2AMC encoding for inference.
In particular, we can still consider the same inner layer and transformation function.
For the outer layer, instead, we need to consider
\begin{itemize}
  \item as semiring, $\mathcal{R}_o^{\mathit{MPE}} = (\mathbb{R}^2 \times {2^{|\probfacts|}}^2, \mathit{max}^4, \mathit{times}^4, (0,0,\probfacts,\probfacts), (1,1,\{\},\{\}))$;
  \item as variables $\probfacts$;
  \item as weight function :
    \begin{equation*}
        w_m(a) =
        \begin{dcases*}
      (p,p,\{a\},\{a\}) & if $a{=}f$ with $p::f$\\
      (1-p,1-p,\{\},\{\}) & if $a{  = }not \ f$ with $p::f$\\ 
      (1,1,\{\},\{\}) & otherwise
        \end{dcases*}
    \end{equation*}
\end{itemize}
where $\mathit{max}^4((x_0,x_1,R_0,R_1),(y_0,y_1,S_0,S_1)) = (z_0,z_1,T_0,T_1)$ with
\begin{itemize}
    \item if $x_0 > y_0$, $z_0 = x_0$; else $z_0 = y_0$;
    \item if $x_1 > y_1$, $z_1 = x_1$; else $z_1 = y_1$;
    \item if $x_0 > y_0$,
        $T_0 = R_0$;
    else if $x_0 < y_0$,
        $T_0 = S_0$;
    else (if $x_0 = y_0$),
        $T_0 = \mathit{min}_>(R_0,S_0)$, where $>$ fixes a total order on $2^{|\probfacts|}$;
    \item if $x_1 > y_1$,
        $T_1 = R_1$;
    else if $x_1 < y_1$,
        $T_1 = S_1$;
    else (if $x_1 = y_1$),
        $T_1 = \mathit{min}_>(R_1,S_1)$, where $>$ fixes a total order on $2^{|\probfacts|}$;
\end{itemize}
and $\mathit{times}^4((x_0,x_1,R_0,R_1),(y_0,y_1,S_0,S_1)) = (x_0 \cdot y_0, x_1 \cdot y_1, R_0 \cup S_0, R_1 \cup S_1)$.
The first two components store, respectively, the value associated with the lower and upper MPE states, while the last two components store the facts selected for such states.

\begin{theorem}
\label{th:commutative_outer_layer_mpe}
$\mathcal{R}_o^{\mathit{MPE}}$ is commutative.
\end{theorem}
\begin{proof}
As before, we have to prove that both $\mathit{max}$ and $\mathit{times}$ are commutative.
The proof for $\mathit{times}$ is easy, as both multiplication and union of sets is commutative.
The proof for $\mathit{max}$ follows a similar path.
This is possible thanks to the total order on the sets of atoms that allows writing $\mathit{min}_>(S_x,S_y) = \mathit{min}_>(S_y,S_x)$, ensuring commutativity.
\end{proof}

\subsection{Encoding MPE as an ASP(Q) Program}
\label{subsec:mpe_asp_aspq}
\ASPQ{} supports weak constraints (both within each subprogram and on the whole \ASPQ{}) to define preferences over quantified answer sets, so it is a natural candidate for modeling MPE in optimal PASP, given its promising results in solving the same task for plain PASP~\citep{DBLP:conf/ijcai/AzzoliniMRR25}.

As shown in the proofs of Theorem~\ref{th:membership_MPE_upper} and~\ref{th:membership_MPE_lower}, the brave and cautious MPE problems can be framed as \ASPQ{} having two quantifiers: the first \textit{guesses} a selection of \pfs{}, while the second \textit{searches} over the optimal \ass{} of the corresponding world for a set that satisfies or not a given query.
In this way, each quantified answer set corresponds to a candidate MPE state.
Then, by expressing preferences among quantified \ass{} by means of global weak constraints, we can select as optimal quantified answer set the candidate MPE state with the maximum probability, thus obtaining a valid MPE state.
This idea is formalized in the following definition.
\begin{definition}\label{def:mpe_in_aspq}
    Let $\optPASP$ be an optimal PASP of the form $(\ASP,\paspfacts,\weakconstraints)$ and $q$ be an atom then:
    \begin{itemize}
        \item []$\qquad\overline{ASPQ}(\optPASP)$ is an \ASPQ of the form $\exists^{st} P_1 \exists^{st} P_2:C:C^\omega$;
        \item []$\qquad\underline{ASPQ}(\optPASP)$ is an \ASPQ of the form $\exists^{st} P_1 \forall^{st} P_2:C:C^\omega$; and
    \end{itemize}
    where $P_1 = \{a;na\mid p::a \in \paspfacts\}$, $P_2 = \ASP$, $C = \{\impl \naf q\}$ and $C^\omega$ is of the form:
    $$
    C^\omega = \left\{\begin{array}{lr}
         :\sim a\ [-log(p)@1,a] & \forall p::a \in \paspfacts  \\
         :\sim na\ [-log(1-p)@1,a] & \forall p::a \in \paspfacts  \\
    \end{array}\right\}
    $$
    with $na$ being fresh atoms not appearing anywhere else.
\end{definition}
To illustrate the \ASPQ{} encoding, we consider the \textit{Optimal Reviewer Assignment} problem from Section~\ref{sec:modelling_examples}.

\paragraph{Brave MPE.}
Suppose we want to compute the brave MPE for the query $\mathit{less\_than\_two}$.
Starting from the optimal PASP presented in Section~\ref{sec:modelling_examples}, namely $\optPASP$, and applying Definition~\ref{def:mpe_in_aspq}, $\overline{ASPQ}(\optPASP)$ is of the form $\exists^{st} P_1 \exists^{st}P_2:C :C^\omega$ shown below:
\begin{lstlisting}[caption={Brave MPE in \ASPQ.}, label={lst:brave_mpe}]
%@exists |\textcolor{olive}{$P_1$}|
reviewer(r1);no_reviewer(r1).  |$\qquad\qquad$|   reviewer(r2);no_reviewer(r2). 
reviewer(r3);no_reviewer(r3).  |$\qquad\qquad$|   reviewer(r4);no_reviewer(r4). 

%@exists |\textcolor{olive}{$P_2$}|
paper(p1). |$\qquad\qquad$| paper(p2).
willing(R,P):- reviewer(R), paper(P).

assign(R,P) :- willing(R,P), not skip(R,P).
skip(R,P) :- willing(R,P), not assign(R,P).

reviewed(P):- assign(R,P).

load(R,2):- assign(R,P1), assign(R,P2), P1!=P2.
load(R,1):- assign(R,P), not load(R,2).
load(R,0):- reviewer(R), not load(R,1), not load(R,2).
less_than_two :- load(R,C), C < 2.

delta(R0,R1,D):- load(R0,L0), load(R1,L1), R0 != R1, L0<=L1, D=L1-L0.
:|$\sim$| assign(R1,P), assign(R2,P), assign(R3,P), R1!=R2, R1!=R3, R2!=R3. [1@2,P]
:|$\sim$| paper(P), not reviewed(P). [1@1,P]
:|$\sim$| delta(R0,R1,D). [D@0,R0,R1]

%@constraint |\textcolor{olive}{$C$}|
:- not less_than_two.

%@global |\textcolor{olive}{$C^\omega$}|
:|$\sim$| reviewer(r1). [-log(0.8)@1,r1] 
:|$\sim$| reviewer(r2). [-log(0.1)@1,r2]
:|$\sim$| reviewer(r3). [-log(0.7)@1,r3]
:|$\sim$| reviewer(r4). [-log(0.2)@1,r4]

:|$\sim$| no_reviewer(r1). [-log(0.2)@1,r1] 
:|$\sim$| no_reviewer(r2). [-log(0.9)@1,r2]
:|$\sim$| no_reviewer(r3). [-log(0.3)@1,r3]
:|$\sim$| no_reviewer(r4). [-log(0.8)@1,r4]
\end{lstlisting}
According to the coherence of \ASPQ{}, a quantified \as{} of $\overline{ASPQ}(\optPASP)$ is an optimal \as{} of $P_1$ (i.e., $M_1\in \optAS(P_1)$) such that by fixing it in the following program, then remaining \ASPQ{} is coherent (i.e., $\exists^{st} P_2\cup \fix{P_1}{M_1}:C:C^{\omega}$ is coherent).
The program $P_1$ contains no weak constraints, so each \as{} is also optimal.
Specifically, an \as{} of $P_1$ corresponds to a possible selection of \pfs{} (i.e., $0.8::reviewer(r1)$, $0.1::reviewer(r2)$, $0.7::reviewer(r3)$, and $0.2::reviewer(r4)$) as each rule of the form $reviewer(r);no\_reviewer(r)$ either selects $r$ as reviewer or not.
Let us first consider the empty selection of reviewers and so $M_1 = \{no\_reviewer(r_1),no\_reviewer(r_2),no\_reviewer(r_3),no\_reviewer(r_4)\} \in \optAS(P_1)$.
We need to verify the coherence of $\exists^{st} P_2\cup\fix{P_1}{M_1}:C:C^\omega$.
We can observe that, since no reviewer is fixed by $M_1$, then no paper can be assigned, and so the only optimal \as{} is $M_2 = M_1 \cup \{paper(p1),paper(p2)\}$.
Thus, the query $less\_than\_two$ is false in $M_2$ and so $M_2$ is such that the program $C$ is incoherent.
Since $M_2$ is the only optimal \as{} of $P_2\cup\fix{P_1}{M_1}$, then $M_1$ cannot be a quantified \as{} of $\overline{ASPQ}(\optPASP)$. 
Let us move on to another selection, for instance the one in which the reviewers $r_1$ and $r_2$ accept to serve as a reviewer and so $M_1 = \{reviewer(r1),reviewer(r2),no\_reviewer(r3),no\_reviewer(r4)\}$.
In this case, if we fix $M_1$ in the program $P_2$ (so we consider optimal \as{} of $P_2\cup\fix{P_1}{M_1}$), then we have three optimal answer sets, namely $M_2^1$, $M_2^2$, and $M_2^3$, where:
\begin{itemize}
    \item $M_2^1$ is such that both papers are assigned to both reviewers;
    \item $M_2^2$ is such that paper $p1$ (resp. $p2$) is assigned to reviewer $r1$ (resp. $p2$); and 
    \item $M_2^3$ is such that paper $p1$ (resp. $p2$) is assigned to reviewer $r2$ (resp. $p1$). 
\end{itemize}
All the papers are covered (i.e., \wcs{} at level $1$ do not fire) and each reviewer has the same number of papers (i.e., \wcs{} at level $0$ do not fire either).
For $M_2^1$ we can note that each reviewer has two papers assigned, and so $less\_than\_two$ is false in $M_2^1$.
Here $M_2^1$ makes the program $C$ incoherent and so it cannot be used to prove that $M_1$ is a quantified \as{}.
Conversely, for $M_2^2$ (resp. $M_2^3$) both reviewers have less than two paper assigned, and so $less\_than\_two$ is true in $M_2^2$ (resp. $M_2^3$). Thus, $M_2^2$ is an optimal \as{} of $P_2\cup\fix{P_1}{M_1}$ that satisfies the constraint in $C$ and so $M_1$ is quantified \as{} of $\overline{ASPQ}(\optPASP)$.

Global weak constraints assign a cost to $M_1$ according to whether the reviewers accepted or not.
Since $reviewer(r1)$ and $reviewer(r2)$ are true in $M_1$, whereas $reviewer(r3)$ and $reviewer(r4)$ are false in $M_1$, then the corresponding weak constraints fires, and so the cost of $M_1$ is $-log(0.8)-log(0.1)-log(0.3)-log(0.8) = -log(P(\sigma))$, with $\sigma=\{reviewer(r1),reviewer(r2)\}$.
At the same time, we can observe that $w_\sigma$ admits an optimal \as{} where paper $p1$ is assigned to reviewer $r1$ and $p2$ is assigned to reviewer $r2$ and so the query $less\_than\_two$ is bravely entailed. 
At this point $\sigma$ is a brave MPE state if its probability $P(\sigma)$ is maximal.
Since the cost of $M_1$ is equal to $-log(P(\sigma))$ and optimal quantified \ass{} are those quantified \ass{} with the minimum cost, then optimal quantified \ass{} are those who maximize the probability of the corresponding selection.
More precisely, an optimal quantified \as{} of $\overline{ASPQ}(\optPASP)$ is $M_1^* = \{reviewer(r1),no\_reviewer(r2),reviewer(r3),no\_reviewer(r4)\}$, which corresponds to the brave MPE state $\sigma = \{reviewer(r1),reviewer(r3)\}$.

\paragraph{Cautious MPE.}
Cautious MPE states can be computed in a similar fashion by computing optimal quantified answer set of $\underline{ASPQ}(\optPASP)$.
Thus, let us assume that we want to compute cautious MPE for $less\_than\_two$, then in this case $\underline{ASPQ}(\optPASP)$ is of the form $\exists^{st} P_1\forall^{st} P_2:C:C^\omega$ where $P_1$, $P_2$, $C$, and $C^\omega$ are the same as in Listing~\ref{lst:brave_mpe}.
Let us now compute a quantified \as{} of $\underline{ASPQ}(\optPASP)$.
Consider $M_1 = \{reviewer(r1),reviewer(r2),no\_reviewer(r3),no\_reviewer(r4)\} \in AS(P_1)$.
From the previous example, we know that by fixing $M_1$ in the program $P_2$ (so we consider optimal \as{} of $P_2\cup\fix{P_1}{M_1}$), we obtain three optimal answer sets, namely $M_2^1$, $M_2^2$, and $M_2^3$, where:
\begin{itemize}
    \item $M_2^1$ is such that both papers are assigned to both reviewers;
    \item $M_2^2$ is such that paper $p1$ (resp. $p2$) is assigned to reviewer $r1$ (resp. $p2$); and 
    \item $M_2^3$ is such that paper $p1$ (resp. $p2$) is assigned to reviewer $r2$ (resp. $p1$). 
\end{itemize}
In this case, program $P_2$ is universally quantified, and therefore the constraint program $C$ must be coherent for every \as{} $M_2^i$, with $1 \leq i \leq 3$.
This condition is not satisfied, as $M_2^1$ assigns all papers (i.e., $p1$ and $p2$) to all selected reviewers (i.e., $r_1$ and $r_2$), thereby violating $C$.
Hence, $M_2^1$ makes $C$ incoherent, and consequently $M_1$ is not a quantified answer set of $\underline{ASPQ}(\optPASP)$.

Conversely, consider the following \as{} of $P_1$ 
$$M_1 = \{reviewer(r1), no\_reviewer(r2), reviewer(r3), reviewer(r4)\}$$ 
corresponding to the selection of reviewers $r1$, $r3$, and $r4$.
By fixing $M_1$ in $P_2$, we obtain six optimal \as{}, each assigning one paper to two of the selected reviewers and the other paper to the remaining reviewer.
In all these cases, each reviewer is assigned exactly one paper, and thus every optimal answer set satisfies the constraint program $C$.
Therefore, $M_1$ is a quantified \as{} of $\underline{ASPQ}(\optPASP)$, and the corresponding selection of reviewers represents a candidate cautious MPE state.

Finally, global weak constraints, also in this case, assign a cost to $M_1$ corresponding to the negative logarithm of the probability of the corresponding selection $\sigma=\{reviewer(r1),reviewer(r3),reviewer(r4)\}$, namely $-\log(0.8) - \log(0.9) - \log(0.7) - \log(0.2) = -log(P(\sigma))$.
Since no better quantified answer set exists, $M_1$ is also optimal.
Note that optimal quantified \ass{} are obtained by minimizing the cost associated with quantified \ass{} and so we are indeed maximizing the log probability of the corresponding selection, and so $\sigma = \{reviewer(r1), reviewer(r3), reviewer(r4)\}$ is a cautious MPE state.

\section{Experimental Evaluation}
\label{sec:experiments}

\paragraph{Implementation.}
The proposed approaches for the inference and MPE problems in optPASP have been implemented on top of existing systems.
More precisely, for the inference problem, we propose two implementations: one that leverages a knowledge-compilation~\citep{DBLP:journals/jair/DarwicheM02} based solver, namely aspmc~\citep{aspmc_aij}, and one based on enumeration.
We call the approaches \textit{aspmc} and \textit{enum}, respectively.
The asmpc tool allows us to specify only the structure of the required semirings, leaving all the remaining parts unchanged.
In a nutshell, aspmc converts the input program into an arithmetic circuit, and then the probability of a query can be computed by traversing this graph and applying the corresponding semiring operations at every node.
The probability values associated to \pfs{} do not influence the execution time, as the size of the compiled programs depends only on the structure of the deterministic part.
For the enumeration-based approach, we developed a tool which converts the input optPASP into an ASP and, for each possible selection (exponentially many), calls an ASP solver, namely clingo~\citep{gebser2019clingo} \added{v 5.8.0}, to compute the optimal answer sets.
Whereas, for the MPE problem, we implemented the proposed encoding in \ASPQ{} and then use the \textsc{casper}\footnote{Note that \textsc{casper} has been recently extended to handle \ASPQ program containing weak constraints and is available at~\url{https://github.com/ndria00/Casper.git}} system~\citep{DBLP:conf/aaai/CuteriMR26} to evaluate the resulting programs.

\paragraph{Dataset Description.}
We performed an experimental evaluation on three different datasets, which extend the application scenarios discussed in Section~\ref{sec:modelling_examples}.
For the 
\textit{viral} datasets (which models the viral marketing problem), we generate instances with an increasing number of people involved in the network. 
Here, the edges (representing friendship relations) are generated according to a Barab\'{a}si Albert network structure (using the NetworkX library~\citep{networkx} with parameter $n$ set to the number of people involved and $m$ set to 2).
For the \textit{paper} dataset (which models the paper assignment problem) we generated instances by iteratively increasing the number of papers $p$ and the number of reviewers $r$.
Lastly, for the \textit{scheduling} dataset (which models the scheduling under no show behavior problem), we generated instances by iteratively increasing the number of exams $e$, people $p$, and resources $r$.
All \pfs{} are assigned random probabilities.
For all instances, we fixed the seed for reproducibility.

The experiments\footnote{Source code and datasets available at: \url{https://drive.google.com/file/d/1NjhLPbD6gqKSI-L9uUHM72QsvHbzLbnc/view?usp=sharing}.}
were executed on a machine running at 3.7 GHz with 32 GB of RAM and a time limit, for each instance, of 3600 seconds (1 hour).
We used the bash command $\mathit{time}$ to extract the execution time and collected the $\mathit{real}$ values.

\subsection{Results}
\label{subsec:results_analysis}

\paragraph{Inference.}
Table~\ref{tab:results_paper_assignments_inference} shows the results for the \textit{paper} dataset.
Here, the approach based on enumeration (\textit{enum}) seem to be effective when the number of articles is small, providing better performance w.r.t. \textit{aspmc} for 2 and 3 papers.
Then, for the remaining cases, \textit{aspmc} can solve significantly more instances.
The largest solvable instance for \textit{enum} is with 6 papers and 5 reviewers while with aspmc it is possible to get up to 10 papers and 7 reviewers.

\begin{table}[tb]
    \centering
    \caption{Results on the \textit{paper} dataset.
    The columns include: 
    the number of articles considered (\textit{\# papers}),
    the number of reviewers associated with the largest solvable instance for a fixed number of papers (\textit{n rev max}),
    the runtime associated with the largest solvable instance for a fixed number of papers (\textit{runtime largest}), and
    the cumulative runtime (i.e., the sum of all runtime for all instances) for a fixed number of papers (\textit{cumulative runtime}), for both \textit{aspmc} and \textit{enum}.
    A dash indicates that none of the instances where solved for the fixed number of papers.
    All times are in seconds.
    }
    \label{tab:results_paper_assignments_inference}
\begin{tabular}{c|c|c|c|c|c|c}
\topline
 & \multicolumn{2}{c|}{\textit{n rev max}} & \multicolumn{2}{c|}{\textit{runtime largest}} & \multicolumn{2}{c}{\textit{cumulative runtime}} \\
\textit{\# papers} & \textit{aspmc} & \textit{enum} & \textit{aspmc} & \textit{enum} & \textit{aspmc} & \textit{enum}\\
\cline{1-7}

2 & 11 & 18 & 740 & 2647 & 1117 & 4091 \\
3 & 9 & 12 & 101 & 1542 & 156 & 1924 \\
4 & 9 & 8 & 450 & 2506 & 555 & 2723 \\
5 & 8 & 6 & 203 & 2143 & 223 & 2217 \\
6 & 8 & 5 & 332 & 3307 & 363 & 3307 \\
7 & 9 & - & 281 & - & 858 &  - \\
8 & 8 & - & 143 & - & 273 & - \\
9 & 8 & - & 357 & - & 629 & - \\
10 & 7  & - & 323 & - & 667 & - 
    \end{tabular}
\end{table}

Table~\ref{tab:scheduling_aspmc_enum} shows the results for the \textit{scheduling} dataset and reports the cumulative runtime obtained by adding up the execution time of all instances with number of resources and exams ranging from 2 to 10, with a fixed number of people.
Here, both approaches can solve all instances.
Then, we further increase the size of the instances to better assess \textit{aspmc} (Table~\ref{tab:scheduling_only_aspmc}).
As for the previous dataset, enum is more effective for small instances.
However, as the size grows, the benefits of KC-based approaches become evident.
The largest number of people that can be handled with \textit{aspmc} is 15.
With 16, it can handle up to 5 resources and 10 exams, requiring approximately 20 minutes.

\begin{table}[tb]
    \centering
    \caption{Results on the \textit{scheduling} dataset. The columns include: the number of considered people (\textit{\# people}) and the cumulative runtime (in seconds) of all instances with a fixed number of people and an increasing number of resources and exams (from 2 to 10). Both solvers can solve all instances.}
    \label{tab:scheduling_aspmc_enum}
    \begin{tabular}{c|c|c}
    \topline
    \textit{\# people} & \textit{aspmc} & \textit{enum} \\
    \cline{1-3}
    2 & 223.35 & 5.33 \\
    3 & 121.86 & 4.88 \\
    4 & 173.86 & 7.95 \\
    5 & 200.26 & 26.94 \\
    6 & 229.58 & 118.13 \\
    7 & 239.54 & 676.83 \\
    8 & 259.19 & 3383.26 \\
    9 & 278.91 & 13120.06 
    \end{tabular}
\end{table}

\begin{table}[tb]
    \centering
    \caption{Results on the \textit{scheduling} dataset. The columns include: the number of considered people (\textit{\# people}) and the cumulative runtime (in seconds) of all instances with a fixed number of people and an increasing number of resources and exams (from 2 to 10). Only \textit{aspmc} can solve these instances (\textit{enum} always timeouts).}
    \label{tab:scheduling_only_aspmc}
    \begin{tabular}{c|c}
    \topline
    \textit{\# people} & \textit{aspmc}  \\
    \cline{1-2}
        10 & 2101.35 \\
        11 & 6323.06 \\
        12 & 13707.62 \\
        13 & 44601.58 \\
        14 & 110555.42 \\
        15 & 223968.87 \\
    \end{tabular}
\end{table}

Lastly, Table~\ref{tab:viralmarketing} shows the results for the \textit{viral} dataset.
Also in this case the approach based on \textit{aspmc} is much more effective w.r.t. enumeration for larger instances.

\begin{table}[t]
    \caption{Results on the \textit{viral} datasets. A dash indicates that the instance cannot be solved.}
    \label{tab:viralmarketing}
    \centering
    \begin{tabular}{c|c|c}
    \topline
        \textit{\# people} & \textit{enum} & \textit{aspmc} \\
        \cline{1-3}
        3 & 0.25 & 4.35 \\
        4 & 27.12 & 3.69 \\
        5 & - & 3.71 \\
        6 & - & 3.92 \\
        7 & - & 5.06 \\
        8 & - & 10.84 \\
        9 & - & 48.85 \\
        10 & - & 338.1
    \end{tabular}
\end{table}

For all datasets, \textit{enum} stops due to timeout, while \textit{aspmc} due to a knowledge compilation error, possibly due to excessive memory requirements.
Overall, enumeration-based approaches are effective for small instances.
However, as the size of the dataset increases, enumeration became unfeasible.
There is a crucial point to consider: \textit{enum} computes only optimal \ass{} while \textit{aspmc} computes \textit{all} answer sets and then only considers optimal ones.
Thus, the number of \ass{} considered in \textit{aspmc} may be much larger than the number considered with \textit{enum}.


\paragraph{MPE.}
For MPE, we focus on \ASPQ{} only.
This is because \textit{enum} enumerates all \ass{}, so the execution time to solve either inference or MPE is the same.
Similar considerations apply to \textit{aspmc} as it consider all \ass{} and then filters them according to the semiring operations.
Furthermore, \ASPQ{} has been already been proven much more effective than \textit{aspmc} to solve MPE in plain PASP~\citep{DBLP:conf/ijcai/AzzoliniMRR25}.

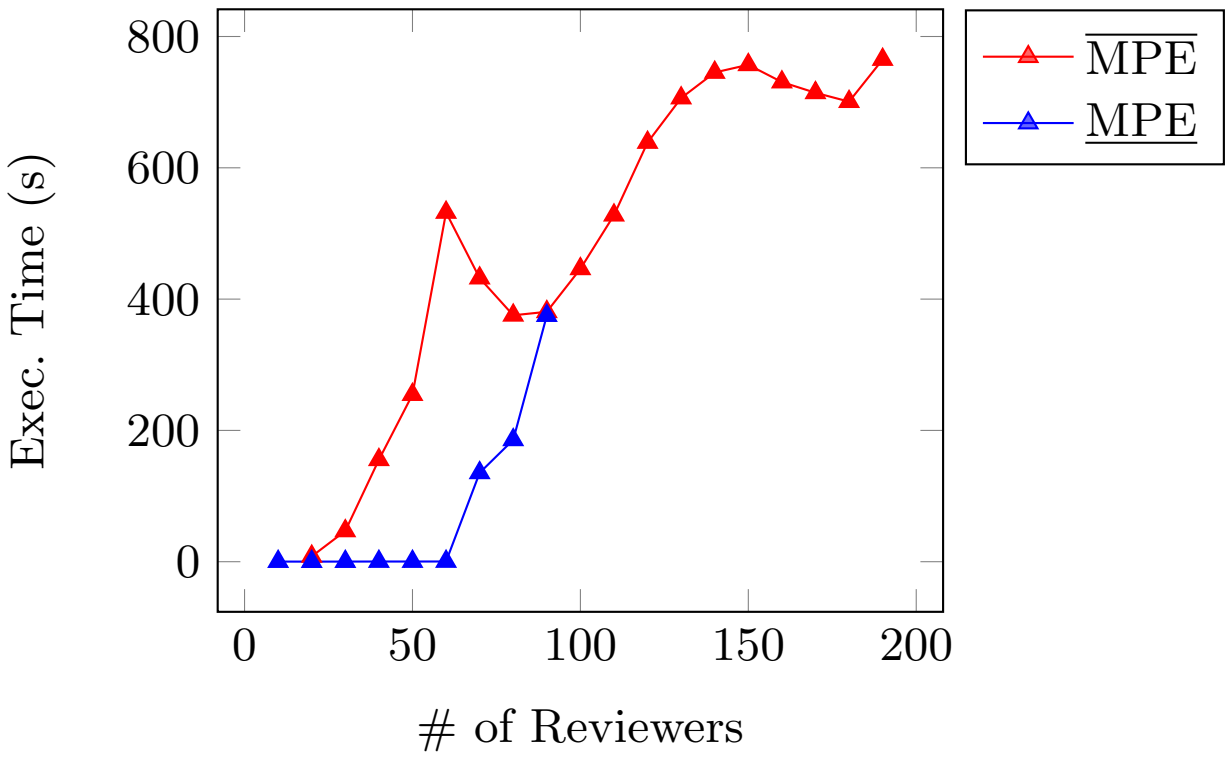
\begin{figure}[tb]
\centering
\begin{tikzpicture}
    \begin{axis}[
        xlabel={\# of Reviewers},
        ylabel={Exec. Time (s)},
        legend pos=outer north east,
        scale=\scaleplot,
        grid style=dashed,
        legend style={fill=white, fill opacity=0.6, draw opacity=1,text opacity=1},
        every axis plot/.append style={thick}
        ]

        \addplot[thin, color = red, mark=triangle*] 
            table [x index=0, y index=1, col sep=comma] 
            {pa_mpe_aspq_upper.txt};
        \addlegendentry{$\uppermpe$}
        \addplot[thin, color = blue, mark=triangle*] 
            table [x index=0, y index=1, col sep=comma] 
            {pa_mpe_aspq_lower.txt};
        \addlegendentry{$\lowermpe$}
    \end{axis}
\end{tikzpicture}
\caption{Results of \ASPQ{} for computing the lower and upper MPE states for the \textit{paper} dataset.}
\label{fig:res_aspq_paper}
\end{figure}
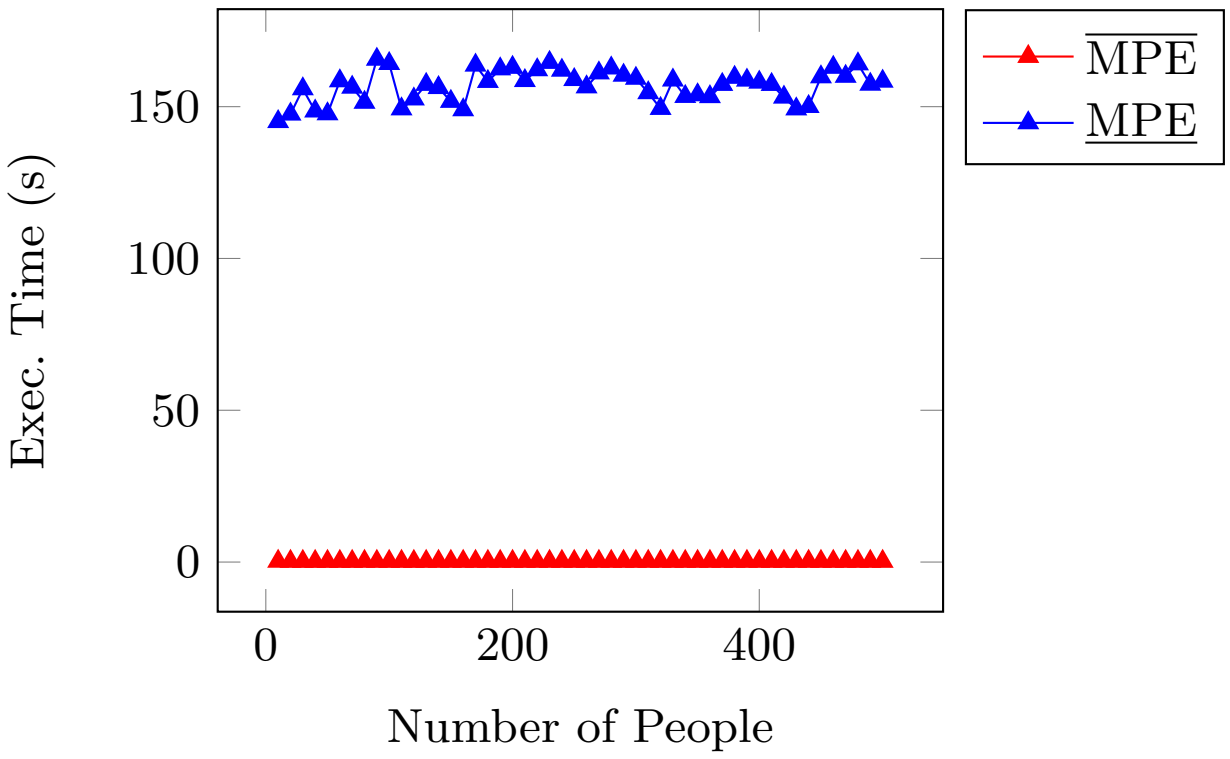
\begin{figure}
    \centering
\begin{tikzpicture}
            \begin{axis}[
                xlabel={Number of People},
                ylabel={Exec. Time (s)},
                legend pos=outer north east,
                scale=\scaleplot,
                grid style=dashed,
                every axis plot/.append style={thick}
                ]            
                \addplot[thin, color = red, mark=triangle*] 
                    table [x index=0, y index=1, col sep=comma] 
                    {vm_mpe_aspq_upper.txt};
                \addlegendentry{$\uppermpe$}
                \addplot[thin, color = blue, mark=triangle*] 
                    table [x index=0, y index=1, col sep=comma] 
                    {vm_mpe_aspq_lower.txt};
                \addlegendentry{$\lowermpe$}
                        \end{axis}
        \end{tikzpicture}
    \caption{Results of \ASPQ{} for computing the lower and upper MPE states for the \textit{viral} dataset.}
    \label{fig:res_aspq_viral}
\end{figure}
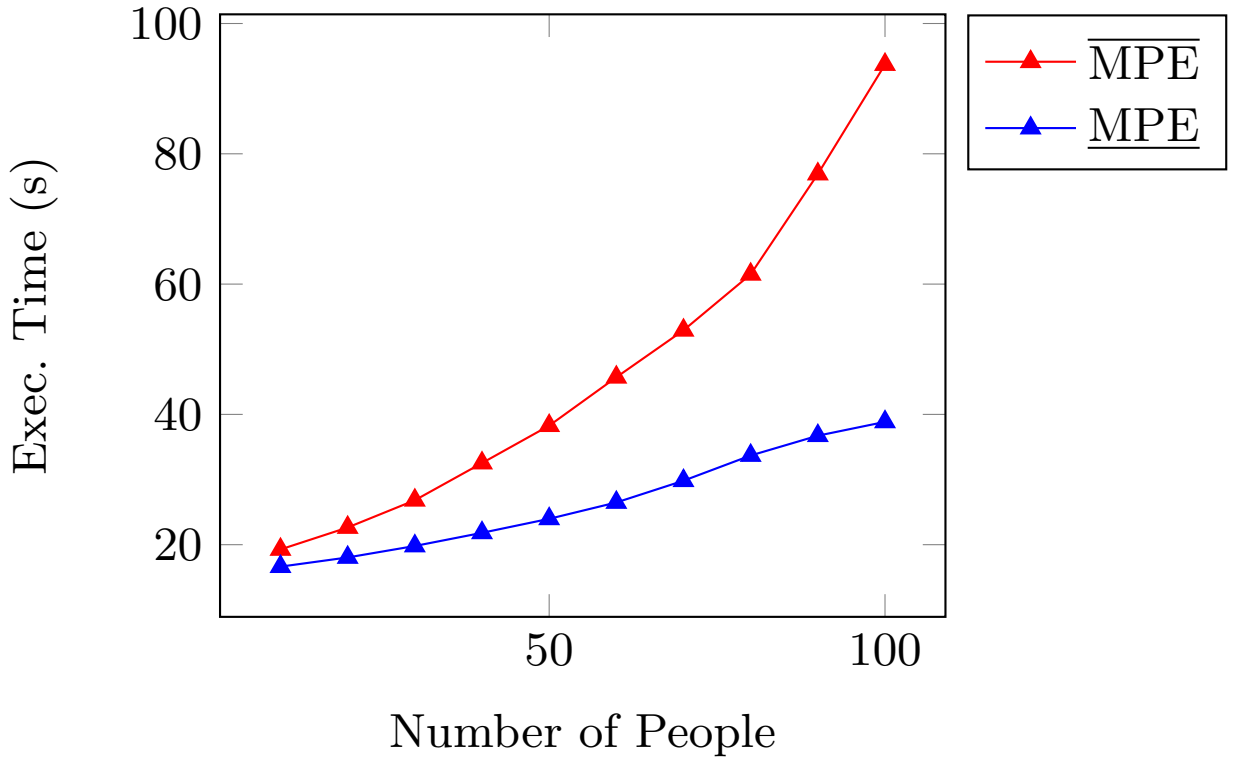
\begin{figure}[tb]
    \centering
        \begin{tikzpicture}
            \begin{axis}[
                xlabel={Number of People},
                ylabel={Exec. Time (s)},
                legend pos=outer north east,
                scale=\scaleplot,
                grid style=dashed,
                legend pos=outer north east,
                every axis plot/.append style={thick}
                ]            
                \addplot[thin, color = red, mark=triangle*] 
                    table [x index=0, y index=1, col sep=comma] 
                    {sns_mpe_aspq_upper.txt};
                \addlegendentry{$\uppermpe$}
                \addplot[thin, color = blue, mark=triangle*] 
                    table [x index=0, y index=1, col sep=comma] 
                    {sns_mpe_aspq_lower.txt};
                \addlegendentry{$\lowermpe$}
                        \end{axis}
        \end{tikzpicture}
    \caption{Results of \ASPQ{} for computing the lower and upper MPE states for the \textit{scheduling} dataset.}
    \label{fig:res_aspq_sched}
\end{figure}

Figures~\ref{fig:res_aspq_paper},~\ref{fig:res_aspq_viral} and~\ref{fig:res_aspq_sched} show the results for computing the MPE with \ASPQ{} for the three datasets.
For \textit{paper} and \textit{scheduling}, the reported results are computed by summing the execution times over all instances with the considered number of reviewers (for \textit{paper}) and people (for \textit{scheduling}).
For \textit{paper}, the we note that, once the number of reviewers is fixed, the number of papers does not seem to influence the execution time.
Thus, the figure reports only the results with 10 articles, as the other are similar.
For \textit{scheduling}, we considered exam and resources both ranging from 10 to 100 with step 10.

For the \textit{paper} dataset, computing the lower MPE state is much slower than computing the upper MPE state, as the largest solvable instance within the time limit is the one with 100 reviewers.
On the contrary, we can compute the upper MPE state for the instance with 200 reviewers within 800 seconds.
For the \textit{viral} dataset, computing the MPE state is almost instantaneous (less than 0.5 seconds) for all instances.
The computation of the lower MPE state, however, requires a bit more time, but still less than 3 minutes.
Lastly, we have a different trend for the \textit{scheduling} dataset, as there the computation of the upper MPE state is slower w.r.t. the computation of the lower MPE state.
Still, in both cases, we can handle the instance with 100 people within 100 seconds for the upper and 40 for the lower.

More in details, for lower MPE, \textsc{casper} directly applies the CEGAR (Counterexample-Guided Abstraction Refinement) approach without any additional translation, since the \ASPQ encoding is alternating (i.e., of the form $\exists^{st} P_1 \forall^{st} P_2 : C : C^\omega$).
In contrast, for upper MPE, \textsc{casper} relies on a preprocessing step that eliminates weak constraints via the translation proposed by~\cite{DBLP:journals/tplp/MazzottaRT24}, as the CEGAR methodology cannot be directly applied to uniform \ASPQ programs (i.e., $\exists^{st} P_1 \exists^{st} P_2 : C : C^\omega$).

These two approaches exhibit different strengths and limitations, which explain the trends observed in the experimental results. On the one hand, the translation required for uniform \ASPQ programs introduces a duplication of the second program within the first one and enforces optimality checks in the second quantifier. As a consequence, the search space increases, since candidate quantified answer sets correspond to pairs $\sigma \cup M$, where $\sigma$ is a selection and $M$ is an answer set of $w_\sigma$. This overhead is particularly evident in the \textit{scheduling} dataset, where \textsc{casper} takes a longer execution time for solving the upper MPE problem compared to lower MPE one. 

On the other hand, for lower MPE, no such translation is not required, and CEGAR can be applied directly. The method explores the space of quantified answer sets by iteratively identifying counterexamples (i.e., optimal answer sets of the second program that invalidate candidate quantified answer set) and refining the original program accordingly. This approach proves to be highly effective on the \textit{scheduling} datasets, while showing some overhead on the \textit{paper} dataset due to the large number of counterexamples that must be discovered before converging to a quantified answer set.

Overall, the results highlight that both upper and lower MPE in optimal PASP are computationally-hard tasks. 
Unlike plain PASP, the two problems share the same computational complexity in the optimal PASP setting, in fact both problems are $\stp$-complete. 
Despite this inherent complexity, \textsc{casper} is very effective in computing the lower and upper MPE state for optimal PASP program, as the majority of the considered instances can still be solved within few minutes.
Lastly, if we compare the results obtained with \ASPQ{} with aspmc, we note that \ASPQ{} is much more efficient.

\section{Related Work}
\label{sec:related}

Related works mainly follows two different directions: alternative semantics for expressing uncertainty within a logic-based language and tasks that can be considered in PASP/PLP frameworks.

\paragraph{Alternative Semantics.}
The credal semantics imposes no preference over the distribution of the probability mass of a world over its \ass{}, introducing probability bounds (lower and upper) for each query.
Here, we specialize this by considering only the optimal \ass{}, according to a set of constraints, but we still keep the two bounds.
The smProbLog~\citep{totis_de_raedt_kimmig_2023} semantics is closely related: here, the probability of a world is uniformly distributed over the \ass{}.
That is, for a world $w$, each of its \as{} has a probability of $\prob(w)/|AS(w)|$ where $\prob(w)$ is computed as in Equation~\ref{eq:prob_world}.
Then, the probability of a query is computed as the sum of the probabilities of the \ass{} where the query is true.
However, smProbLog only considers plain PASP (without \wcs{}) for worlds.
Extending this semantics considering \wcs{} is an interesting direction for future work.

Both CS and smProbLog require each world to be coherent (i.e., with at least one \as{}).
If this does not hold, there is a loss of probability mass.
The L-credal semantics~\citep{DBLP:conf/kr/RochaC22} tries to overcome this by introducing a ``undefined'' truth value.
Another possibility is to normalize the probability~\citep{azzolini2025solving}, i.e., dividing the probability of the query by 1 minus the sum of the probabilities of incoherent worlds.
Another possibility is to ``repair'' the program~\citep{eiter2016generalized,ceylan2016complexity}.

Other semantics for PASP include
LPMLN~\citep{lee_talsania_wang_2017} that extends ASP with weighted rules and atoms.
Furthermore, there is the possibility of handling infinite weights that represent hard constraints.
In LPMLN, there is no concept of world.
Rather, each \as{} is associated with a weight, and the probability of a query is the sum of the probabilities (obtained by dividing the weight of an \as{} by the total weight) of the \ass{} containing the query.
Another approach is P-log~\citep{baral2009probabilistic}, where it is possible to define random attributes and a probability distribution over the answer sets.
The relationship between LPMLN an P-log has been extensively explored in~\citep{lee2017lpmln}.

If worlds are plain answer set programs composed of normal rules only and every such program has a unique stable model (which also coincides with the unique well-founded model~\citep{van1991well}), we fall into the Probabilistic Logic Programming (PLP) field~\citep{DBLP:conf/ijcai/RaedtKT07,Rig23-BKaddress}.

\paragraph{Related Tasks.}
In this paper, we focus on inference and computation of the MPE state in optPASP.
MPE has also been studied for plain PASP and PLP~\citep{DBLP:conf/ilp/ShterionovRVKMJ14,DBLP:conf/ijcai/AzzoliniMRR25}.
In PLP, the task can be solved with a plain ASP, by converting \pfs{} into choice rules and considering log-probabilities.
However, for PASP, saturation~\citep{DBLP:journals/amai/EiterG95,DBLP:journals/csur/DantsinEGV01} or more advanced approaches such as Answer Set Programming with Quantifiers~\citep{DBLP:journals/tplp/AmendolaRT19,DBLP:journals/tplp/MazzottaRT24} are needed.
MPE is closely related to Maximum-a-Posteriori, which has been studied for both PLP~\citep{bellodi_alberti_riguzzi_zese_2020} and PASP~\citep{azzolini2025algebraic}.
In brief, in MAP some \pfs{} are denoted as \textit{query} \pfs{} and the goal is to compute a subset of such facts (called MAP state) such that the probability of the given evidence is maximized.
This generalizes MPE, as in MPE all \pfs{} are considered as query \pfs{}.
Computing the MAP state is harder than solving MPE~\cite{cozman2020complexity} and standard approaches are based on knowledge compilation.

Decision theory (DT) tasks can be framed in the PLP and PASP frameworks~\citep{DBLP:conf/aaai/BroeckTOR10,azzolini2025solving}.
Here, the PLP/PASP is extended with utility attributes, which specify costs/utilities associated with some of the atoms of the program, and decision atoms, which can be selected or not. 
Each selection of decision atoms identifies a PLP/PASP which is associated with an expected cost by considering the probability and the utilities of the different literals.
The goal is to find the set of decision atoms that maximizes the expected utility.
This is different w.r.t. inference in optPASP, as in DT the maximization is on the outer level (over the decision atoms) while in optPASP, \wcs{} impose an optimality criterion over \ass{} (so, in the inner level) and then these are considered for the computation of the probability.
Similarly, the MPE also differs, as it aims to find the optimal solution over all \pfs{}, not only a subset of it (as may happen in DT with decision atoms).
Furthermore, the result of DT is a tuple composed of a real number and a set of atoms, while for MPE we have a probability.

Lastly, another related task is abduction in PLP~\citep{AzzBelFerRigZes21-IJAR-IJ}: here, a PLP is extended with abducible facts and integrity constraints.
Each selection of abducible facts identifies a PLP.
Given a query $q$, the goal is to find the minimal set (where minimality is considered under set-inclusion) of abducible facts that maximizes the probability of $q$.
Extending all these tasks to the optPASP framework is also an interesting direction for future work.

The inference, MPE, MAP, and decision theory problems, for both PLP and PASP, have been all formalized in the AMC framework.
For PLP, inference and MPE can be effectively encoded with AMC~\citep{10.1016/j.jal.2016.11.031,DBLP:conf/ijcai/AzzoliniMRR25}, while MAP and decision theory with 2AMC~\citep{DBLP:journals/tplp/KieselTK22}.
For PASP, inference and MPE can be effectively encoded with 2AMC~\citep{DBLP:conf/ijcai/AzzoliniMRR25} while MAP and decision theory with 3AMC~\citep{azzolini2025solving,azzolini2025algebraic}.
Overall, in this paper we frame inference in optPASP within the AMC spectrum, thus providing a connection with existing works.

\section{Conclusion}
\label{sec:conclusion}
In this paper, we propose \textit{optimal} Probabilistic Answer Set Programming (optPASP), an extension of Probabilistic Answer Set Programming that incorporates \wcs{}, a standard and powerful ASP construct for modeling preferences.
After presenting the syntax and semantics of optPASP, we illustrated its modeling capabilities through three real-world–inspired applications: the paper assignment problem, the viral marketing problem, and scheduling under no-show behavior. These case studies demonstrate the expressiveness and practical potential of our framework.
On the theoretical side, we studied the computational complexity of the main task associated with optASP: 
inference and most probable explanation (MPE).
On the practical side, we encoded these tasks as instances of Algebraic Model Counting, and we also proposed an implementation in Answer Set Programming with Quantifiers. 
We evaluated the resulting implementations on real-world-inspired benchmark problems, demonstrating the practical viability of our approach.

\medskip

Several directions for future work can be envisaged. On the one hand, it would be interesting to extend other semantics for PASP to accommodate \wcs{}, thereby broadening the applicability of optimization in probabilistic settings. On the other hand, further investigation of additional reasoning tasks in optPASP, such as Maximum-a-Posteriori (MAP) inference, represents a natural and promising direction.

\section*{Competing Interests}
Competing interests: The authors declare none.

\bibliographystyle{acmtrans}
\bibliography{bibliography}

\end{document}